\documentclass[useAMS,usenatbib]{mn2e}
\usepackage{stmaryrd}
\usepackage{amssymb}
\usepackage{amsmath}
\usepackage{bbm}
\usepackage{graphicx}

\def\gsim{\;\lower4pt\hbox{${\buildrel\displaystyle >\over\sim}$}\;}
\def\lsim{\;\lower4pt\hbox{${\buildrel\displaystyle <\over\sim}$}\;}
\def\grls{\;\lower4pt\hbox{${\buildrel\displaystyle >\over <}$}\;}

\title[Polytropic Shock Flows in H \Rmnum{2} Regions]
{Self-Similar Polytropic Champagne Flows in H \Rmnum{2} Regions}
\author[R.-Y. Hu and Y.-Q. Lou]{Ren-Yu Hu$^{1}$\thanks{E-mail address:
hu-ry07@mails.tsinghua.edu.cn (RYH) and louyq@tsinghua.edu.cn
(Y-QL)} and Yu-Qing Lou$^{1}$$^{,2}$$^{,3}$\footnotemark[1]\\
$^{1}$Physics Department and Tsinghua Centre for Astrophysics (THCA),
  Tsinghua University, Beijing 100084, China\\
$^{2}$Department of Astronomy and Astrophysics, The University of
Chicago, 5640 South Ellis Avenue, Chicago, IL 60637, USA\\
$^{3}$National Astronomical Observatories, Chinese Academy of
Science, A20, Datun Road, Beijing, 100012, China}
\begin{document}

\date{Accepted 2008 ?? ??. Received 2008 ?? ??; in original form 2008 ?? ??}

\makeatletter
\newcommand{\rmnum}[1]{\romannumeral #1}
\newcommand{\Rmnum}[1]{\expandafter\@slowromancap\romannumeral #1@}
\makeatother

\pagerange{\pageref{firstpage}--\pageref{lastpage}} \pubyear{2008}

\maketitle

\label{firstpage}

\begin{abstract}
We explore large-scale hydrodynamics of H \Rmnum{2} regions for
various self-similar shock flows of a polytropic gas cloud under
self-gravity and with quasi-spherical symmetry. We formulate cloud
dynamics by invoking
specific entropy conservation along streamlines and obtain global
self-similar ``champagne flows" for a conventional polytropic gas
with shocks as a subclass. Molecular cloud cores are ionized and
heated to high temperatures after the onset of nuclear burning of
a central protostar. We model subsequent evolutionary processes in
several ways and construct possible self-similar shock flow
solutions. We may neglect the mass and gravity of the central
protostar. The ionization and heating of the surrounding medium
drive outflows in the inner cloud core and a shock travels
outwards,
leading to the so-called ``champagne phase" with an expanding
outer cloud envelope. Complementarily, we also consider the
expansion of a central cavity around the centre. As the inner
cloud expands plausibly due to powerful stellar winds, a cavity
(i.e., `void' or `bubble') can be created around the centre, and
when the cavity becomes sufficiently large, one may neglect the
gravity of the central protostar. We thus present self-similar
shock solutions for ``champagne flows" with an expanding central
void. We compare our solutions with isothermal solutions and find
that the generalization to the polytropic regime brings about
significant differences of the gas dynamics, especially for cases
of $n<1$, where $n$ is a key scaling index in the self-similar
transformation. We also compare our global polytropic self-similar
solutions with numerical simulations on the expansion of H
\Rmnum{2} regions. We further explore other possible dynamic
evolutions of H \Rmnum{2} regions after the initiation of nuclear
burning of the central protostar, for example asymptotic inflows
or contractions far from the cloud centre and the ongoing infall
around a central protostar. In particular, it is possible to use
the downstream free-fall solution with shocks to describe the
dynamic evolution of H \Rmnum{2} regions shortly after the
nascence of the central protostar. We also give an analysis on the
invariant form of self-similar polytropic flows by ignoring
self-gravity.
\end{abstract}


\section{Introduction}


The compact (scale sizes of $\sim 0.1-1$ pc) and ultracompact (UC;
scale sizes $\lsim 0.15$ pc) H \Rmnum{2} regions are associated with
massive OB stars \citep[e.g.,][]{Habing}.
The UC H \Rmnum{2} region stage ($\lsim 10^5$ yr) stands for a
substantial part of the relatively short main-sequence lifetimes of
OB stars.
Several radio surveys (e.g., Wood \& Churchwell 1989; Fish 1993;
Kurtz, Churchwell \& Wood 1994) observed expansions of luminous H
\Rmnum{2} regions with shock signatures.
%
%
``Champagne flow" models \citep[e.g.,][]{Tenorio1979, TT, York}
successfully explain the expansion of H \Rmnum{2} regions by
considering a protostar formed in a cloud core, photonionizing and
heating the cloud, as well as driving a shock that accelerates the
ionized gas to expand rapidly. Observations tend to support
``champagne flow" models, such as Lumsden \& Hoare (1999) for UC H
\Rmnum{2} regions G 29.96-0.02 and \citet{Barriault} for Compact H
\Rmnum{2} region Sh 2-158. Champagne flows in clouds of larger
scales have also been identified, such as \citet{Foster} for the
dense
Galactic H \Rmnum{2} region G84.9+0.5 and Maheswar et al. (2007) for
the classical
H \Rmnum{2} region S236 in the cluster of OB stars NGC 1893.
%

\citet{Tenorio1979} classified champagne flows into cases R and
cases D. For the case R, the ionization front (IF) created by the
emergence of the central massive protostar rapidly breaks out from
the dense cloud and leaves the gas behind it fully ionized. For
the case D, the ionization front is `trapped' inside a cloud, and
produces an expanding H \Rmnum{2} region within a cloud.
In the formation phase of H \Rmnum{2} regions, whether the IF is
R-type or D-type depends on the initial grain opacity, the
ionizing flux, and the initial density and the size of a cloud
\citep{FTB}.
In the expansion phase, H \Rmnum{2} regions with an initial mass
density profile $\rho\propto r^{-l}$ and $l>3/2$ are `density
bounded', where $\rho$ is the mass density and $r$ is the radius
(see, e.g., Osterbrock 1989; Franco et al. 1990).
\footnote{In Franco et al. (1990),
the power-law index of the mass density profile is denoted by $w$
instead of $l$ as adopted here to avoid notational confusions.} It
is also possible that the D-type IF changes to a weak R-type IF
when $l>3/2$. In such cases, the fully ionized cloud begins to
expand and an outgoing shock forms. This is referred to as the
``champagne phase"
(Bodenheimer et al. 1979). Notably if a shock front encounters a
steep negative density gradient, for example, the edge of a cloud,
asymmetric ``champagne flows" may occur, as observed by Lumsden \&
Hoare (1999). According to the VLA survey by Wood \& Churchwell
(1989), 16\% of H \Rmnum{2} regions bear cometary appearance.
\citet{Arthur} numerically simulated ``cometary champagne flows".
If $l<3/2$, the H \Rmnum{2} region is `ionization bounded', e.g.,
the ultraviolet radiation is trapped within a finite radius
\citep{Oster}. In such cases, the ionized region should expand as
$t^{4/(7-2l)}$, driving a shock that would accelerate the ambient
medium into a thin shell (Franco et al. 1990).
In this paper, we focus on the champagne phase of a cloud assumed
to be `density bounded', implying $l>3/2$. We also assume that the
cloud is fully ionized shortly after the onset of nuclear burning
of a central massive protostar.

\citet{Shu} (also Tsai \& Hsu 1995) investigated isothermal
``champagne flows" under spherical symmetry in the self-similar
framework. By neglecting the gravity of the central massive
protostar and assuming that a cloud initially stays at rest and
gets heated by the luminous massive protostar to a uniform high
temperature, one may obtain self-similar expansion solutions
connected with isothermal outflows with shocks; this corresponds
to a ``champagne flow" in a highly idealized setting. The initial
isothermal mass density profile scales as $\rho\propto r^{-2}$. In
addition, for molecular clouds with other possible initial mass
density profiles, \citet{Shu} also ignored the self-gravity of a
cloud completely and proposed another self-similar transformation,
referred to as the `invariant form', and obtained solutions for
cases with an initial mass density profile $\rho\propto r^{-l}$,
where the power-law index $l$ is not necessarily equal to 2. In
general, it is not realistic to suppose molecular clouds to be
isothermal in many astrophysical situations. One specific example
of cloud temperature measurement of a UC H \Rmnum{2} region NGC
6334F undergoing a ``champagne flow" reveals a conspicuous
temperature gradient from the centre to the edge
\citep[e.g.,][]{DeBuizer}.

Energy sources and plasma coolings in molecular clouds are not
completely known. We then approximate the energy equation by a
general polytropic equation of state $p=\kappa(r,\ t)\
\rho^{\gamma}$, where $p$ is the thermal gas pressure, $\gamma$ is
the polytropic index and the proportional coefficient $\kappa(r,\
t)$ (related to specific entropy) depends on radius $r$ and time
$t$ in general. Setting $\kappa$ as a global constant, the
equation of state simply becomes a conventional polytropic one. By
adjusting $\gamma$, we may model various situations of H \Rmnum{2}
regions in molecular clouds. For example, for $\gamma=1$ and a
constant $\kappa$, our solutions reduce to isothermal ones. Since
``champagne flows" in a polytropic molecular cloud has not been
studied, we would generalize the isothermal analyses of
\citet{Shu} and of Tsai \& Hsu (1995) to a polytropic description
of self-similar ``champagne flows". We shall provide the basic
formulation with the most general polytropic equation of state
(i.e., specific entropy conservation along streamlines; Wang \&
Lou 2008) and present global ``champagne flow" solutions with
shocks for a conventional polytropic gas.

\citet{Shu}
introduced the Bondi-Parker radius as a measure for the effective
distance of the central gravity. The Bondi-Parker radius is defined
by
$r_{\rm BP}=GM_*/(2a^2)\ ,$
where $M_*$ is the mass of the central gravity source and $a$ is
the sound speed of the surrounding medium, which is a constant for
an isothermal cloud. The mass originally residing within a radius
$r_0$ is dumped into the star during the star formation. After the
star formation, as the surrounding cloud becomes much hotter with
a higher sound speed, the Bondi-Parker radius becomes much less
than $r_0$. Therefore for the gas in $r>r_0$, the gravity of the
central massive star may be neglected. This reasoning naturally
leads to the possible existence of a cavity around the centre of a
molecular cloud, which we refer to as `void' or `bubble'. At
$t=0$, the void boundary is at $r=r_0$.
For an expanding cloud, the central void also expands. Indeed, a
stellar wind drives also a principal shock and is capable of
sweeping the surrounding ionized gas into an expanding shell.
Wood \& Churchwell (1989) identified central cavities in the shell
or cometary UC H \Rmnum{2} regions, which are thought to be
supported by stellar wind and radiation pressures in their survey.
Lumsden \& Hoare (1999) suggested a ``champagne flow" surrounding
a hot stellar wind bubble to interpret observations of G
29.96-0.02. \citet{Comeron} numerically simulated the dynamic
evolution of wind-driven H \Rmnum{2} regions with strong density
gradients and found that features of classical champagne model
are not substantially changed, except that the compression of the
swept-up matter would, rapidly and particularly in densest cases,
lead to the trapping of the IF and inhibit the champagne phase.
Therefore, the dynamic evolution of void expansion represents an
important physical aspect of ``champagne flows". Recently, Lou \&
Hu (2008, in preparation) explore self-similar solutions for voids
in a more general context.
In this paper, we construct self-similar solutions for ``champagne
flows" with central voids in self-similar expansion. The inclusion
of a central void not only makes our model more realistic, but
also allows us to take into account stellar wind bubbles.

We outline the model formulation of a general polytropic gas and
present self-similar asymptotic solutions in section 2 and
construct global solutions of ``champagne flows" in section 3.
Section 4 provides solutions of self-similar ``champagne flows"
with an expanding central void. In section 5, we discuss
behaviours and astrophysical applications of our novel solutions,
and suggest other plausible forms of H \Rmnum{2} regions.
Details of an invariant form of self-similar solutions in a
conventional polytropic gas with the self-gravity ignored are
summarized in Appendix A.

\section[]{Self-Similar Polytropic Flows}

\subsection[]{General Polytropic Formulation}

Dynamic evolution of a quasi-spherical general polytropic gas
under self-gravity can be described by nonlinear hydrodynamic
equations in spherical polar coordinates $(r,\ \theta,\ \phi)$,
\begin{equation}
\frac{\partial \rho}{\partial
t}+\frac{1}{r^2}\frac{\partial}{\partial r}(r^2\rho u)=0\ ,
\label{equ1}
\end{equation}
\begin{equation}
\frac{\partial M}{\partial t}+u \frac{\partial M}{\partial r}=0\ ,
\label{equ2}
\end{equation}
\begin{equation}
\frac{\partial M}{\partial r}=4\pi r^2\rho\ ,\label{equ3}
\end{equation}
\begin{equation}
\rho \bigg(\frac{\partial u}{\partial t}+u \frac{\partial
u}{\partial r}\bigg)=-\frac{\partial p}{\partial r}-\frac{GM
\rho}{r^2}\ ,\label{equ4}
\end{equation}
\begin{equation}
p=\kappa(r,\ t)\rho^{\gamma}\ ,\label{equ5}
\end{equation}
where $\rho(r,\ t)$ is the mass density, $u(r,\ t)$ is the bulk
gas radial flow velocity, $M(r,\ t)$ is the enclosed mass within
$r$ at time $t$, $p$ is the thermal pressure, $G=6.67\times
10^{-8}$ dyne cm$^2$ g$^{-2}$ is the gravity constant. Equations
(\ref{equ1}), (\ref{equ2}), (\ref{equ3}) describe the mass
conservation, equation (\ref{equ4}) is the radial momentum
equation and equation (\ref{equ5}) is the general polytropic
equation of state, in which $\gamma$ is the polytropic index and
the coefficient $\kappa$ directly related to the `specific
entropy' depends on $r$ and $t$. For a conventional polytropic
gas, $\kappa$ is a global constant in space and time. More
generally, we require the conservation of `specific entropy' along
streamlines, namely
\begin{equation}
\bigg(\frac{\partial}{\partial t}+u\frac{\partial}{\partial
r}\bigg)\bigg(\ln\frac{p}{\rho^{\gamma}}\bigg)=0\ . \label{equ5a}
\end{equation}
This set of equations is the same as those of \citet{WangLou08}
but without a completely random magnetic field.

To reduce the nonlinear partial differential equations (PDEs) to
ordinary differential equations (ODEs) for self-similar flows, we
introduce the following transformation
\begin{eqnarray}
r=k^{1/2} t^n x\ ,\qquad u=k^{1/2} t^{n-1} v\ ,\qquad
\rho=\frac{\alpha}{4\pi G
t^2}\ ,\nonumber\\
p=\frac{k t^{2n-4}}{4\pi G}\beta\ ,\quad\qquad M=\frac{k^{3/2}
t^{3n-2} m}{(3n-2)G}\ ,\label{equ6}
\end{eqnarray}
where $x$ is a dimensionless independent self-similar variable,
$k$ is a dimensional parameter related to the polytropic sound
speed making $x$ dimensionless, $v(x)$, $\alpha(x)$, $\beta(x)$,
$m(x)$ are dimensionless reduced dependent variables of $x$ only,
and $n$ is a key scaling index which controls the relation between
$r$ and $x$ as well as various scalings of reduced dependent
variables. We refer to $v(x)$, $ \alpha(x)$, $\beta(x)$, and
$m(x)$ as the reduced radial flow speed, mass density, thermal
pressure, and enclosed mass, respectively. Transformation
(\ref{equ6}) is identical with that of Lou \& Wang (2006).

By performing self-similar transformation (\ref{equ6}) in
equations (\ref{equ1})$-$(\ref{equ5a}) and introducing parameter
$q\equiv 2(n+\gamma-2)/(3n-2)$, we obtain two integral relations
\begin{equation}
m=\alpha x^2 (nx-v)\ ,\label{equ7}
\end{equation}
\begin{equation}
\beta=\alpha^\gamma m^q\ ,\label{equ8}
\end{equation}
where there is no loss of generality to set the proportional
coefficient (i.e., an integration constant) equal to unity in
integral (\ref{equ8}) for $q\neq 2/3$ or $\gamma\neq
4/3$~\citep{WangLou08}. The special case of $\gamma=4/3$
corresponds to a relativistically hot gas as studied by Goldreich
\& Weber (1980) and Lou \& Cao (2008).
By setting $q=0$, the general polytropic formulation reduces to
the conventional polytropic case of a global constant $\kappa$
(e.g., Suto \& Silk 1988; Lou \& Gao 2006; Lou \& Wang 2006; Lou,
Jiang \& Jin 2008) with $n+\gamma=2$. According to expression
$M(r,t)$ for the enclosed mass in transformation (\ref{equ6}), we
require $3n-2>0$ and $nx-v>0$ to ensure a positive enclosed mass.
The inequality $3n-2>0$ will reappear later for a class of
asymptotic solutions at large $x$. By equation (\ref{equ7}), we
emphasize that for $nx-v=0$ at a certain $x^*$, the enclosed mass
within $x^*$ becomes zero; we refer to this as a central void and
$x^*$ is the independent similarity variable marking the void
boundary which expands with time $t$ in a self-similar manner.

Combining all reduced equations above, we readily derive two coupled
nonlinear ODEs for $\alpha'$ and $v'$ as
\begin{equation}
{\cal X}(x,\alpha,v)\alpha'={\cal A}(x,\alpha,v)\ ,\ \ {\cal X}(x,
\alpha, v)v'={\cal V}(x,\alpha, v)\ ,\label{equ9a}
\end{equation}
where functionals ${\cal X}$, ${\cal A}$ and ${\cal V}$ are defined
by
\begin{eqnarray}
\!\!\!\!\! & \!\!\!\!\! {\cal
X}(x,\alpha,v)\equiv\big[2-n+(3n-2)q/2\big]\alpha^{1-n+3nq/2}
\nonumber\\& \qquad\times x^{2q}
(nx-v)^q-(nx-v)^2\ ,\nonumber\\
&{\cal A}(x,\alpha,v)\equiv 2\frac{x-v}{x}\alpha
\bigg[q\alpha^{1-n+3nq/2} x^{2q} (nx-v)^{q-1}
\nonumber\\&+(nx-v)\bigg]-\alpha\bigg[(n-1)v
+\frac{nx-v}{3n-2}\alpha\nonumber\\&+ q\alpha^{1-n+3nq/2}x^{2q-1}
(nx-v)^{q-1}(3nx-2v)\bigg]\ ,\nonumber\\& {\cal V}(x,\alpha,
v)\equiv 2 \frac{x-v}{x}\alpha\bigg(2-n+\frac{3n}{2}q\bigg)
\alpha^{-n+3nq/2} x^{2q}\nonumber\\& \times(nx-v)^q
-(nx-v)\bigg[(n-1)v+\frac{nx-v}{3n-2}\alpha \nonumber\\&
+q\alpha^{1-n+3nq/2} x^{2q-1} (nx-v)^{q-1}(3nx-2v)\bigg]\ .
\label{equ10a}
\end{eqnarray}
For a conventional polytropic gas of constant $\kappa$ with
$n+\gamma=2$, we simply set $q=0$ in equation (\ref{equ10a}) to
derive
\begin{eqnarray}
\!\!\!\!\!\!\!\!\!\!
\frac{\alpha'}{\alpha^2}=\frac{(n-1)v+\frac{(nx-v)\alpha}{(3n-2)}
-2(x-v)(nx-v)/x}{\alpha(nx-v)^2-\gamma\alpha^{\gamma}}\ ,\label{equ9}\\
v'=\frac{(n-1)\alpha
v(nx-v)+\frac{(nx-v)^2}{(3n-2)}\alpha^2-2\gamma\alpha^{\gamma}
(x-v)/x}{\alpha(nx-v)^2-\gamma\alpha^{\gamma}}\ .\label{equ10}
\end{eqnarray}
Up to this point, our basic self-similar hydrodynamic formulation is
the same as that of Lou \& Wang (2006) and of \citet{WangLou08}
without a random magnetic field.


For energy conservation, we define the energy density $\epsilon$
and the energy flux density ${\cal J}$ as follows,
\begin{equation}
\epsilon=\frac{\rho u^2}{2}-\frac{GM\rho}{r}+\frac{i}{2}p\ ,
\label{Eden}
\end{equation}
\begin{equation}
{\cal J}=\rho
u\bigg(\frac{u^2}{2}-\frac{GM}{r}+\frac{i}{2}\frac{\gamma
p}{\rho}\bigg)\ ,\label{Eflux}
\end{equation}
where $\epsilon$ is the energy density, ${\cal J}$ is the energy
flux density and $i$ is the degree of freedom of an individual gas
particle. The three terms in expressions (\ref{Eden}) and
(\ref{Eflux}) correspond to densities of the kinetic energy, the
gravitational energy and the internal energy, respectively. With
equations $(\ref{equ1})-(\ref{equ5})$ and a globally constant
$\kappa$, we derive
\begin{equation}
\frac{\partial\epsilon}{\partial
t}+\frac{1}{r^2}\frac{\partial}{\partial r}(r^2{\cal J})={\cal
P}\equiv u\frac{\partial p}{\partial
r}\bigg[\frac{i}{2}(\gamma-1)-1\bigg] \label{Econ}
\end{equation}
for energy conservation, where ${\cal P}$ represents the net
energy input. If the gas expands adiabatically or
$\gamma=(i+2)/i$, then ${\cal P}=0$. Whether the gas locally gains
or loses energy depends not only on the difference between
$\gamma$ and $(i+2)/i$, but also on the signs of $\partial
p/\partial r$ and $u$.

\subsection[]{Self-Similar Solutions}

An exact globally static solution, known as the singular polytropic
sphere (SPS) takes the following form of
\begin{equation}
v=0\ , \quad \alpha=\bigg[\frac{n^{2-q}}{2(2-n)
(3n-2)}\bigg]^{-1/(n-3nq/2)}x^{-2/n}\ .\label{SPS}
\end{equation}
This is a straightforward generalization of the singular
isothermal sphere (SIS; e.g., Shu 1977) and of the SPS for a
conventional polytropic gas with $q=0$ (Lou \& Wang 2006; Lou \&
Gao 2006). For a general SPS here, the mass density profile scales
as $\rho\propto r^{-2/n}$, independent of $q$ parameter.

For large $x$, the asymptotic flow behaviour is
\begin{eqnarray}
\!\!\!\!\!\!
v=\bigg[-\frac{nA}{(3n-2)}+2(2-n)
n^{q-1}A^{1-n+3nq/2}\bigg]x^{1-2/n}\quad\nonumber\\
+Bx^{1-1/n}+\cdots\ ,\  \label{equ15}\\
\alpha=Ax^{-2/n}+\cdots\
,\qquad\qquad\qquad\qquad\qquad\qquad\qquad\ \ \ \ \label{equ16}
\end{eqnarray}
where $A>0$ and $B$ are two constant parameters. As $x\rightarrow
+\infty$, the mass density profile scales the same way as SPS
(\ref{SPS}) above, independent of $q$ parameter to the leading
order. Since $x\rightarrow +\infty$ means $t\rightarrow 0^{+}$,
the asymptotic solution at large $x$ is thus equivalent to initial
conditions of the system at a finite $r$. Coefficients $A$ and $B$
are referred to as the mass and velocity parameters, respectively.
A global solution with asymptotic behaviour (\ref{equ15}) and
(\ref{equ16}) represents a fluid whose density profile scales
initially similar to that of a SPS. Thus, by varying the scaling
index $n$ or the polytropic index $\gamma$, we are able to model
the initial density profile with an index $l=-2/n$. Furthermore,
as a physical requirement for plausible similarity solutions of
our polytropic flow, both $v(x)$ and $\alpha(x)$ should remain
finite or tend to zero at large $x$. Hence, in cases of $2/3<n<1$,
corresponding to $2<l<3$,
$A$ and $B$ are fairly arbitrary, while in cases of $1\leqslant
n<2$ (general polytropic), corresponding to $1<l\leqslant2$, $B$
should vanish for a finite radial flow velocity at large $x$.

In the regime of $x\rightarrow0^+$, there exists a free-fall
asymptotic solutions for which the gravity is virtually the only
force in action, and the radial velocity and the mass density
profile both diverge in the limit of $x\rightarrow0^+$. To the
leading order, the free-fall asymptotic solution takes the form of
\begin{equation}
\alpha(x)=\bigg[\frac{(3n-2)m(0)}{2x^3}\bigg]^{1/2}\ ,
\label{freefall1}
\end{equation}
\begin{equation}
v(x)=-\bigg[\frac{2m(0)}{(3n-2)x}\bigg]^{1/2}, \label{freefall2}
\end{equation}
where constant $m(0)$ represents an increasing central point mass.
Such solutions were first found by \citet{b1} in the isothermal
case, and were generalized to the conventional polytropic
case~\citep[Cheng 1978;][]{b33, b11} and to the general polytropic
case with a random magnetic field by \citet{WangLou08}. The
asymptotic form does not depend on $q$. The validity of such
solutions requires $n>2/3$ and $\gamma<5/3$; the last inequality
appears as a result of comparing various terms in series
expansions.

Another exact global solution, known as the Einstein-de Sitter
(EdS) solution, exists in two cases of $q=0$ and $q=2/3$. The EdS
solution for a conventional polytropic gas of $q=0$ reads as
\begin{equation}
v=\frac{2}{3}x\ ,\qquad \alpha=\frac{2}{3}\ ,\qquad
m=\frac{2(n-2/3)}{3}x^3\ ,\label{equ11}
\end{equation}
(Lou \& Wang 2006); for $q=2/3$ and thus $\gamma=4/3$, this global
EdS solution without and with a random magnetic field takes a
slightly different form (Lou \& Cao 2008). With shocks, EdS
solutions can be used to construct polytropic ``champagne flows"
with various upstream dynamics.

From now on, we focus on the conventional polytropic case of $q=0$
with $n+\gamma=2$. To simulate ``champagne flows" in H \Rmnum{2}
regions in star-forming clouds, we solve coupled nonlinear ODEs
(\ref{equ9}) and (\ref{equ10}) subject to the inner boundary
conditions at $x=0$, namely,
\begin{equation}
\alpha=\alpha_0\ ,\qquad\qquad v=0\ ,\qquad\ \label{equ12}
\end{equation}
where $\alpha_0$ is a constant. A series expansion yields the LP
type of asymptotic solutions at small $x$ in the form of
\begin{eqnarray}
v=\frac{2}{3}x-\frac{\alpha_0^{(1-\gamma)}}
{15\gamma}\bigg(\alpha_0-\frac{2}{3}\bigg)
\bigg(n-\frac{2}{3}\bigg)x^3+\cdots\ ,
\label{equ13}\\
\alpha=\alpha_0-\frac{\alpha_0^{(2-\gamma)}}
{6\gamma}\bigg(\alpha_0-\frac{2}{3}\bigg)x^2+\cdots\ .\
\qquad\qquad \label{equ14}
\end{eqnarray}
The isothermal counterpart of this polytropic series solution was
obtained earlier \citep[][Hunter 1977; Shu et al.
2002]{Lar1,Lar2,Pen1,Pen2} with $n=1$ and $\gamma=1$. Such LP type
solutions may be utilized to construct champagne flows, if we
ignore the central protostar as an approximation and assume the
surrounding gas to be initially static \citep[e.g.,][]{Shu}.
Physically, as a result of the gravity of the central protostar,
the gas infall towards the very central region may not stop, even
when the protostar starts to shine and the `champagne phase
expansion' has occurred in the outer cloud envelope due to the
photoionization and ultraviolet (UV) heating. Thus with central
free falls (\ref{freefall1}) and (\ref{freefall2}) as the
downstream side of a shock, we can also construct global solutions
for the dynamics of H \Rmnum{2} regions surrounding a nascent
central massive protostar which involves free-fall materials.
We can take LP type or EdS solutions as the downstream side of a
shock and model classical champagne flows for a conventional
polytropic gas. We shall come to the possible scenario of inner
free-fall solutions with an outer champagne flow.

With the restriction $n+\gamma=2$ for a conventional polytropic
gas, the initial density profile with the scaling index $n$ is
directly linked to the polytropic index $\gamma$, depending on the
energy exchange process in the gas (see eq \ref{Econ}). For a
general polytropic gas with $q\neq 0$ in contrast, the scaling
index $n$ and polytropic index $\gamma$ can be independently
specified; in particular, we can have $\gamma>1$ and $n\geq 1$
(this is impossible for a conventional polytropic gas).
On the other hand, the initial mass density profile is affected by
the star formation or other energetic processes before $t=0$.
Hence for $l=-2/n$, we postulate that the energy exchange process
remains largely unchanged after the protostar formation at $t=0$.
This is plausible because the ionization front travels relatively
fast to large distances in a cloud during the ``champagne flow"
phase.
Our general polytropic model allows inequality $2/3<n<2$,
corresponding to $1<l<3$, which covers so far the entire range of
initial mass density profiles of H \Rmnum{2} regions. This range
of mass density distribution has been obtained from radio
observations of cloud fragments and isolated dark clouds
\citep[e.g.,][]{Arquilla, Myers}.

Franco et al. (1990) reveals that the initial mass density profile
index $3/2<l<3$ (i.e., $2/3<n<4/3$ in a polytropic model) leads to
``champagne flows" in clouds with weak shocks; and $l>3$
corresponds to a ``champagne flow" with strong and accelerating
shocks. We then require parameter $n$ within the range
$2/3<n<4/3$, as we assume the cloud is `density bounded'. With
such range of parameter, we provide solutions of ``champagne flow"
with shocks.
%
%
%
%
We emphasize in particular that there is one more degree of
freedom to choose the velocity parameter $B$ in asymptotic
solution (\ref{equ15}) and (\ref{equ16}) in polytropic cases with
$2/3<n<1$ (i.e. $2<l<3$) than in the isothermal case. This leads
to major differences of the polytropic champagne flow solutions we
find as compared with those isothermal solutions of Shu et al.
(2002) and Tsai \& Hsu (1995). \citet{Franco} have recently argued
from the radio continuum spectra that UC H \Rmnum{2} regions have
initial density gradients with $2\leqslant l\leqslant3$; so we
would consider primarily the $n$ range of $2/3<n<1$ or $2<l<3$. In
summary, molecular clouds with $3/2<l<3$ have ``champagne flows"
in self-similar manner, while those clouds with $l\geqslant3$ have
``champagne flows" without similarity. For clouds with $2<l<3$,
there is more than one parameter to specify initial dynamic flows.

Finally, we need to include shocks at proper places in LP type
solutions, EdS solutions or free-fall solutions to match with
appropriate asymptotic solutions at large $x$ to determine
relevant coefficients. The shock jump conditions between the
downstream and upstream variables are determined by the mass
conservation, the radial momentum conservation and the energy
conservation. With these three equations for shock conditions, one
can determine upstream self-similar variables $(x_{s{\rm u}},\
\alpha_{\rm u},\ v_{\rm u})$ uniquely from downstream self-similar
variables $(x_{s{\rm d}},\ \alpha_{\rm d},\ v_{\rm d})$ or vice
versa. The two subscripts ${\rm d}$ and ${\rm u}$ here denote the
downstream and upstream variables, respectively. Detailed
formulation and procedure of self-similar shocks can be found in
Section 5 of Lou \& Wang (2006). All solutions in this paper are
obtained by solving coupled nonlinear ODEs (\ref{equ9}) and
(\ref{equ10}) for a conventional polytropic gas with $n+\gamma=2$.


\section[]{Polytropic Champagne Flows}

In cases of $n<1$,
there is a range of shock positions (or speeds) for a specified
downstream solution with a fixed density at the centre $\alpha_0$
at $x=0$, corresponding to different asymptotic flow behaviours at
large $x$ on the upstream side. We now discuss two situations:
first, for cases with a fixed value of $\alpha_0$, we adjust shock
positions for a specified LP type solution and observe the
relation between shock positions and asymptotic flow behaviours at
large $x$. Secondly, for cases with a fixed shock position, we
alter the value of $\alpha_0$ and examine the variation of
upstream conditions.
It is expected to set certain limits on relevant parameters for
polytropic ``champagne flows" to exist. As a series of examples,
we choose the scaling index $n=0.9$. Numerical explorations have
also been performed for cases of $n=0.7$ and $n=0.8$ and the
results are qualitatively similar.

\subsection[]{Cases with a Fixed $\alpha_0$ Value}

\begin{figure}
 \includegraphics[width=0.5\textwidth]{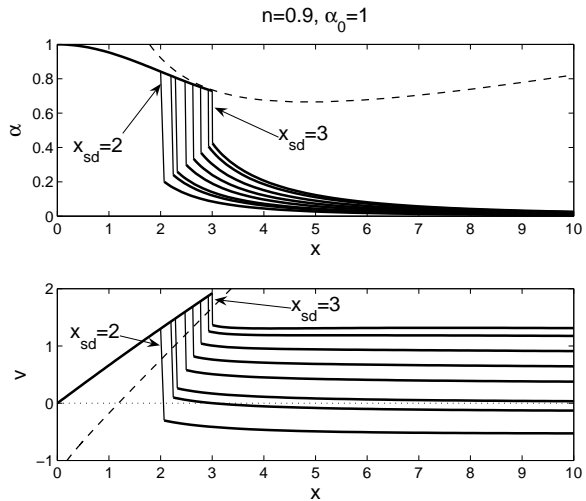}
 \caption{The reduced mass density $\alpha(x)$ (top) and the reduced
 radial flow velocity $v(x)$ (bottom) for global ``champagne flow"
 solutions in cases with $n=0.9$ (thus $\gamma=1.1$) and $\alpha_0=1$.
 In both panels the dashed curve represents the sonic critical curve and in the bottom panel the dotted line is $v=0$.
 The downstream solution is connected with various upstream solutions
 with light solid curves. The downstream solution is integrated
 numerically from $x\rightarrow 0^{+}$ with a LP type solution. In both
 panels, the upstream solutions from top to bottom correspond to
 $x_{s{\rm d}}$=3, 2.923, 2.773, 2.623, 2.473, 2.287, 2.2 and 2 respectively. We note that the upstream
 solution with $x_{s{\rm d}}=2.287$ corresponds
 to a breeze. Relevant parameters are summarized in Table \ref{tab1}.}
 \label{Fig1}
\end{figure}

With a fixed value of $\alpha_0$ for $x\rightarrow 0^{+}$, one can
uniquely determine a LP type solution by a standard numerical
integration. Such a LP type solution will encounter the sonic
critical curve at a certain $x_{\rm max}$ uniquely corresponding
to $\alpha_0$. It is natural to consider possible hydrodynamic
shock positions $x_{s{\rm d}}<x_{\rm max}$ on the downstream side
of shock front. One such example of $n=0.9$ and $\alpha_0=1$ is
shown in Figure \ref{Fig1} with relevant parameters summarized in
Table \ref{tab1}. To model ``champagne flows" with outward
expansions at large radii in molecular clouds, we provide the
following analysis. In principle, there are two other conditions
giving rise to two minima of $x_{s{\rm d}}$ in order to obtain an
outflow in the upstream side of a shock. First, our extensive
numerical explorations reveal that there exists a $x_{\rm min1}$
and for $x_{s{\rm d}}<x_{\rm min1}$, the upstream shock position
$x_{s{\rm u}}$ becomes complex by the shock conditions. This only
happens when we attempt to obtain upstream variables from
downstream variables. For a real $x_{s{\rm u}}$, downstream
variables $(x_{s{\rm d}},\ \alpha_{\rm d},\ v_{\rm d})$ should
satisfy
\begin{equation}
(1-\gamma)\alpha_{\rm d}^{\gamma-1}+2(nx_{s{\rm d}}-v_{\rm
d})^2>0\ .\label{xmin}
\end{equation}
For $\gamma< 1$ (unphysical) and $\gamma=1$, inequality
(\ref{xmin}) is readily satisfied; for $\gamma>1$ or $n<1$, this
condition does not always hold. Algebraic manipulations give the
downstream Mach number in the shock reference framework ${\cal
M}_{\rm d}$ as
\begin{equation}
{\cal M}_{\rm d}^2=\frac{(nx_{s{\rm d}}-v_{\rm d})^2}{\gamma
\alpha_{\rm d}^{\gamma-1}}\ .\label{Math}
\end{equation}
Inequality (\ref{xmin}) imposed on a subsonic downstream Mach
number is $1>{\cal M}_{\rm d}^2>(\gamma-1)/(2\gamma)$. The
downstream Mach number and the upstream Mach number is related by
\begin{equation}
{\cal M}_{\rm d}^2=\frac{2+(\gamma-1){\cal M}_{\rm
u}^2}{2\gamma{\cal M}_{\rm u}^2-(\gamma-1)}\ .
\end{equation}
This relation was provided by \citet{LouCao} for a
relativistically hot gas and is proven valid in our model
consideration. With the possible range of upstream Mach number
$1<{\cal M}_{\rm u}^2<+\infty$, we then have the limit on ${\cal
M}_{\rm d}$ shown above. As we integrate LP or EdS solutions from
$x=0$ with $\alpha=\alpha_0$ and $v=0$, solutions do not satisfy
inequality (\ref{xmin}) when $x$ remains sufficiently small
and $x_{\rm min1}$ is the minimum value of $x$ satisfying
(\ref{xmin}). This value of $x_{\rm min1}$ is uniquely determined by
the $\alpha_0$ value. Therefore, for a fixed LP type solution around
small $x$, polytropic shocks can be constructed with a downstream
shock in the range of $x_{\rm min1}<x_{s{\rm d}}<x_{\rm max}$, and
across such a shock, the LP type solution at small $x$ can be
matched with different asymptotic flows at large $x$.

Systematic numerical explorations reveal that the upstream velocity
increases monotonically with the increase of $x_{s{\rm d}}$, as
shown by the variation trend of the $B$ parameter (see Table
\ref{tab1}). There is thus another critical value imposed on
$x_{s{\rm d}}$, denoted by $x_{\rm min2}$. For $x_{s{\rm d}}>x_{\rm
min2}$, the upstream solution matches to an asymptotic solution at
large $x$ in the form of (\ref{equ15}) and (\ref{equ16}) with $B>0$,
referred to as an outflow. Complementarily with $x_{s{\rm d}}<x_{\rm
min2}$, the upstream solution matches to an asymptotic solution with
$B<0$, referred to as an inflow. As $B$ varies continuously and
monotonically with $x_{s{\rm d}}$, for $x_{s{\rm d}}=x_{\rm min2}$
the upstream solution corresponds to an asymptotic solution with
$B=0$, which describes a breeze or a contraction in association with
``champagne flows". According to asymptotic expressions
(\ref{equ15}) and (\ref{equ16}) with $q=0$, the breeze or
contraction correspond to slow outward or inward flows. To obtain a
breeze, we need a mass parameter
\begin{equation}
A<A_s\equiv \bigg[\frac{n^2}{2\gamma(3n-2)}\bigg]^{-1/n}\ .
\end{equation}
For the specific case of $A=A_s$, either a breeze or a contraction
reduces to SPS solution (\ref{SPS}). With $n<1$, there are three
possibilities in general. First, $x_{\rm min1}>x_{\rm min2}$ for the
allowed range to construct shocks, the upstream solutions always
correspond to outflows.
Secondly, $x_{\rm min1}<x_{\rm min2}$ for the allowed range to
construct shocks, it is possible to obtain outflows, inflows and
breezes or contractions for the upstream side.
Thirdly, if any of $x_{\rm min1}$ or $x_{\rm min2}$ exceeds $x_{\rm
max}$, a global champagne flow is not allowed. For the isothermal
case of $n=1$,
we may set $B$ equal to zero; asymptotic breezes or contractions on
the upstream side are allowed, and the only allowed value of the
downstream shock position is $x_{s{\rm d}}=x_{\rm min2}$.
\citet{Shu} indicated that the isothermal shock position is uniquely
determined by the value of $\alpha_0$, which is consistent with our
more general analysis here. The unique shock position found by
\citet{Shu} corresponds to the $x_{\rm min2}$ above.
%

The conventional scenario for ``champagne flows" would require the
entire fluid to expand outward. Numerical simulations on ``champagne
flows" \citep[e.g.,][]{TT1} assume that at $t=0^{+}$ the central
star is formed and the surrounding cloud is initially at rest. For
$t>0^+$, the fluid is photoionized and heated by the ultraviolet
radiation from the central star and expands. In this scenario, we
would require an expanding upstream flow in order to model
``champagne flows". However, since solutions with an asymptotic
inflow or contraction as upstream part may also exist, the outer
part of H \Rmnum{2} regions can also have inward velocities. In
fact, during the star formation even some time after the star
formation, the surrounding cloud may continue to collapse towards
the centre \citep[e.g.,][]{b32}. With the core nuclear burning of
the central protostar, the surrounding gas is ionized and heated,
and the inner part of the fluid starts to expand, while the outer
part continues to fall inwards. Those solutions of the LP type as
the downstream side of a shock and an asymptotic inflow or
contraction for the upstream side correspond to this scenario just
described; and such global solutions are referred to as Inner Shock
Expansions in a Collapsing Envelope (ISECE).


For the case of $n=0.9$ and $\alpha_0=1$, we have determined
$x_{\rm min1}=0.95$, $x_{\rm min2}=2.287$ and $x_{\rm max}=3$; the
shock range $2.287<x_{s{\rm d}}<3$ gives sensible classical
polytropic ``champagne flow" solutions and $0.95<x_{s{\rm
d}}<2.287$ gives the ISECE solutions as shown in Figure
\ref{Fig1}. For $n=0.9$, we have $A_s=2.042$; with $x_{s{\rm
d}}=x_{\rm min2}$ and $A=1.536<A_s$, the asymptotic solution is a
breeze and should be considered as a classical champagne flow.
Given other parameters the same, the situation of $A>A_s$ would
give rise to an asymptotic contraction. The shock location and
shock speed can be determined once $x_{s{\rm d}}$ is specified.
The dimensionless shock position in the self-similar variable
represents the shock strength and velocity in dimensional form.
The shock velocity reads $dr_s/dt=nk_{\rm d}^{1/2}x_{s{\rm
d}}t^{n-1}$. The outgoing shock slightly decelerates (for $n$
slightly less than 1) and the shock velocity is proportional to
$x_{s{\rm d}}$.

As shown in Figure \ref{Fig1} and Table \ref{tab1}, with
increasing $x_{s{\rm d}}$, the upstream variables at the shock
front $v_{\rm u}$ and $\alpha_{\rm u}$ increase, and the two
parameters $A$ and $B$ of asymptotic solutions (\ref{equ15}) and
(\ref{equ16}) also increase. This is a fairly common feature,
observed in all other cases that we have studied numerically.
Different shock positions match with different asymptotic
solutions at large $x$. Once $A$ and $B$ are specified, the shock
position $x_{s{\rm d}}$ is uniquely determined. As $A$ is the mass
parameter and $B$ is the velocity parameter, $x_{s{\rm d}}$,
proportional to the shock velocity and strength, is determined not
only by initial mass density but also by the initial motion. This
differs from the isothermal case
when $B=0$ with the mass parameter $A$ determining the shock
behaviour. We expect a faster and stronger shock with a higher
initial speed.

We further identify two sub-types of such ISECE solutions: (i) the
upstream side has an outward velocity near the shock and an
asymptotic inward velocity far from the centre (e.g., solution
with $x_{s{\rm d}}=2.2$ in Figure \ref{Fig1}); (ii) the upstream
side has an inward velocity everywhere (e.g., solution with
$x_{s{\rm d}}=2$ in Figure \ref{Fig1}). For the type (i)
solutions, there is a stagnation point $x_{\rm stg}$ where the
radial flow velocity vanishes; for the solution with $x_{s{\rm
d}}=2.2$, $x_{\rm stg}\sim3$. With self-similar transformation
(\ref{equ6}), this stagnation point $r_{\rm
stg}=k^{1/2}t^{n}x_{\rm stg}$ travels outward with time in a
self-similar manner.

For ISECE solutions, we envision that such solutions correspond to
the situation where a star starts to burn, ionizing and heating
the surrounding medium as the gas falling continues. The gas
infall and collapse are indispensable in star formation. If the
nascent star ionizes the whole residual gas sufficiently fast, the
outer gas may still possess an inward momentum. A champagne shock
runs into the infall gas, deposits outward momentum and
accelerates the outer gas. If a shock is sufficiently strong, we
expect type (i) solutions, e.g., the gas immediately outside a
shock flows outward, and the stagnation point travels outward
proportional to the self-similar expansion of the shock. If a
shock is sufficiently weak, we expect type (ii) solutions, e.g.,
the gas flows inward outside the shock front. This ISECE
scenario is expected to occur in certain H \Rmnum{2} regions
(e.g., Shen \& Lou 2004; \citet{b32}).

\begin{table*}
\begin{center}
  \caption{Data Parameters of Global Polytropic ``Champagne Flow" Solutions in Cases with
 $n=0.9$ and $\alpha_0=1$}
  \begin{tabular}{|l|l|l|l|l|l|l|l|}
  \hline
  \hline
   $A$ & $B$ & $x_{s{\rm d}}$ & $\alpha_{\rm d}$ & $v_{\rm d}$
   & $x_{s{\rm u}}$ & $\alpha_{\rm u}$ & $v_{\rm u}$\\
   \hline
      0.9902 & $-0.7699$ & 2 & 0.8416 & 1.3019 & 2.0693 & 0.2005 &
   -0.3012\\
   1.3501 & $-0.2267$ & 2.2 & 0.8172 & 1.4266 & 2.2446 & 0.2404 &
   0.1004\\
1.5357 & 0& 2.2869&  0.8068&  1.4805 &  2.3234&  0.2586&  0.2599\\
2.0109 & 0.4767  &2.4732 & 0.7848  &1.5956  &2.4964  &0.2993  &0.5789\\
2.474 &0.8577 & 2.6232  &0.7678  &1.688   &2.6387  &0.3336  &0.8172\\
3.0224 &1.2416 &2.7731 &0.7516  &1.7803  &2.7832  &0.369   &1.0425\\
3.6696& 1.6323 &2.9231 &0.7363 &1.8728 &2.9292  &0.4054  &1.2568
\\4.0455 &1.8367 &3   &0.729 &1.9204
&3.0047 &0.4243 &1.3629\\
  \hline
\end{tabular}\label{tab1}
\end{center}
\end{table*}

\subsection[]{Cases of a Fixed Dimensionless Shock Position}

\begin{figure}
 \includegraphics[width=0.5\textwidth]{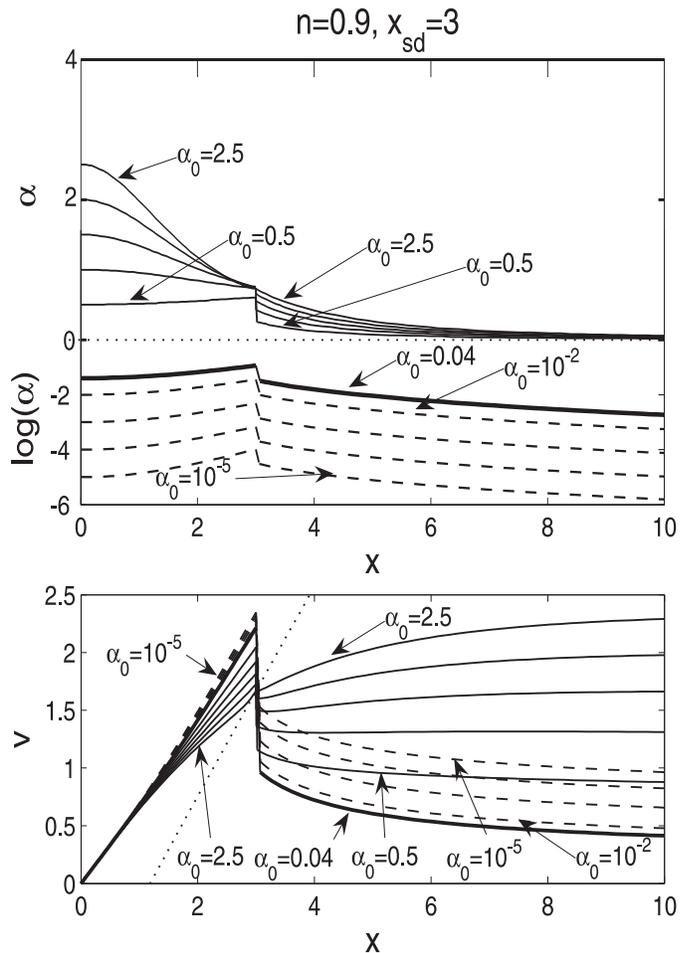}
 \caption{The reduced mass density $\alpha(x)$ (top) and the
 reduced radial flow velocity $v(x)$ (bottom) for global polytropic
 ``champagne flow" solutions in cases $n=0.9$ (thus $\gamma=1.1$)
 and $x_{s{\rm d}}=3$. In the top panel, the horizontal dotted line
 stands for $\alpha=0$, separating the top panel into two
 parts; the vertical scales in these parts are different.
 In the upper part $\alpha(x)$ is presented linearly, while in the lower part
 $\log[\alpha(x)]$ is presented. In the bottom panel, the dotted
 line is the sonic critical curve. Shocks at $x\sim 3$
 as the discontinuity in solutions are shown. The
 solid curves on the downstream side of shocks are LP
 type solutions with $\alpha_0$=0.5, 1, 1.5, 2 and 2.5 (from bottom
 to top in the top panel and from top to bottom in the bottom panel).
 The solid curves on the right show respectively the corresponding
 upstream solutions approaching different asymptotic solutions
 at large $x$ (from bottom to top in the top panel and from bottom
 to top in the bottom panel). The dashed curves on the
 downstream side of shocks are LP type solutions with $\alpha_0=10^{-5}$,
 $10^{-4}$, $10^{-3}$, and $10^{-2}$ (from bottom to top in the top panel
 and from top to bottom in the bottom panel). The dashed curves on the
 right show respectively the corresponding upstream solutions
 approaching different asymptotic solutions at large $x$ (from bottom to
 top in the top panel and from top to bottom in the bottom panel). The
 thick solid curve in both panels represents the solution with
 $\alpha_0=0.04$, which has the lowest upstream speed and the minimum
 $B$ at large $x$. Relevant parameters are summarized in Table \ref{tab2}.}
 \label{Fig2}
\end{figure}

The variation of LP type solutions with different $\alpha_0$ and
the same $x_{s{\rm d}}$ is shown in Figure \ref{Fig2}. For
$\alpha_0>2/3$, $\alpha$ decreases with increasing $x$ while for
$\alpha_0<2/3$, $\alpha$ increases with increasing $x$. With a
larger $\alpha_0$ the LP type solution encounters the sonic
critical curve earlier; therefore $\alpha_0$ cannot be too large,
otherwise the LP type solution encounters the sonic critical curve
before reaching the preset shock position $x_{s{\rm d}}$. One case
of $n=0.9$ and $x_{s{\rm d}}=3$ is shown in Figure \ref{Fig2} with
parameters given in Table \ref{tab2}. Here, we have $\alpha_{s{\rm
d}}<2.5$.

As shown in Figure \ref{Fig2} and Table \ref{tab2}, with the
increase of $\alpha_0$, the upstream conditions at the shock front
$\alpha_{\rm u}$ and the mass parameter $A$ of the asymptotic
solution increases, but $v_{\rm u}$ and $B$ do not have a steady
trend. As $\alpha_0$ increases from $10^{-5}$ to $0.04$, $v_{\rm u}$
and $B$ decrease, while as $\alpha_0$ increases from 0.04 to 2.5,
$v_{\rm u}$ and $B$ increase (see Table \ref{tab2}). This shows that
$v_{\rm u}$ and $B$ are correlated.
The minimum values of $v_{\rm u}$ and $B$ are associated with
$\alpha_0=0.04$. Here we show that with a prefixed downstream shock
position $x_{s{\rm d}}$, it is possible that neither breeze nor
contraction solutions are allowed. With a $x_{s{\rm d}}$ as large as
3, all the asymptotic upstream solutions correspond to outflows.
In the isothermal analysis of \citet{Shu}, the mass parameter $A$
tends to zero with $\alpha_0\rightarrow 0^{+}$. In polytropic cases,
we obtain similar results.
According to series expansions (\ref{equ13}) and (\ref{equ14}), if
setting $\alpha_0=0$ exactly, the integration gives a trivial
solution of $\alpha=0$.

Based on numerical explorations for $\alpha_0\rightarrow 0^{+}$ in
isothermal cases, \citet{Shu} argued that the self-gravity may be
neglected for cases with small central density, and developed
another self-similar transformation, viz., the so-called invariant
form, in order to model the initial mass density profile other than
that of a SIS (i.e., $\rho\propto r^{-2}$). The initial mass density
profile $\rho\propto r^{-l}$, where index $l$ does not necessarily
equal to 2, can be described by the invariant form when the
self-gravity is ignored.
We perform a similar reduction with the invariant self-similar
transformation for a conventional polytropic gas without the
self-gravity (see Appendix A), and show that with $n\neq 1$ (i.e.,
non-isothermal cases), the power index $l$ must be equal to $2/n$
for a self-similar form. This mass density profile with a scaling
index $l=2/n$ is the same as that in a SPS and in asymptotic
solutions at large $x$. In summary, the freedom of choosing $l$ in
the invariant form disappears for non-isothermal cases. From
another perspective, since the index $n$ ranges in $2/3<n<4/3$,
self-similar polytropic champagne flows can model the initial mass
density profile with $3/2<l<3$. In other words, the objective to
model the initial density profile other than $l=2$ is naturally
fulfilled, without the necessity of dropping the self-gravity. The
clear advantage of our polytropic approach is that the
self-gravity is included in the model consideration. Therefore, to
apply our polytropic solutions, $\alpha_0\rightarrow 0^{+}$ is no
longer required. From transformation (\ref{equ6}), parameter
$\alpha_0$ is tightly linked with the central mass density
$\rho_0$ and timescale $t$, and should have different values in
different situations. Therefore polytropic champagne flow
solutions are adaptable to a much wider range of astrophysical
cloud systems. Moreover, the $\alpha_0\rightarrow 0^{+}$ cases in
our polytropic framework can be approximated by a central ``void"
as discussed in the following section. With a central ``void", we
are able to neglect the gravity of the central region where the
density is sufficiently low and still consider the self-gravity of
the outer more dense gas medium.



\begin{table*}
\begin{center}
  \caption{Data Parameters of Global Polytropic ``Champagne Flow"
 Solutions in Cases with $n=0.9$ and $x_{s{\rm d}}=3$.
 }
  \begin{tabular}{|l|l|l|l|l|l|l|l|l|}
  \hline
  \hline
  $\alpha_0$ & $A$ & $B$ & $x_{s{\rm d}}$ & $\alpha_{\rm d}$ & $v_{\rm d}$
  & $x_{s{\rm u}}$ & $\alpha_{\rm u}$ &  $v_{\rm u}$\\
   \hline
   0.00001& 0.0002532& 1.145 & 3 & 0.0001 & 2.3469 & 3.0609 &
   0.000030184&
   1.5294\\
0.0001  &0.001735  &0.9499 & 3& 0.0007&  2.3152 & 3.064 &0.0001993&
1.3959\\ 0.001   &0.01243  &0.7205 & 3  & 0.0049 &2.2821 &3.0693&
0.001372 &1.2353\\ 0.01&    0.0912  &0.4713  &3 &0.0358 & 2.2449
&3.0746 &0.009752 &1.0536\\ 0.04 &0.3045 & 0.3899 & 3& 0.1154 &
2.2135 &3.0715 &0.03192& 0.9641\\ 0.5& 2.4053  &1.1453 &3& 0.6048
&2.0444 &3.0197 &0.2556 &1.1567\\ 1 &4.0455& 1.8367 & 3 & 0.729  &
1.9204 &3.0047 &0.4244& 1.3629
\\1.5& 5.5009& 2.406& 3   &0.7512  &1.8166 & 3.008 &  0.5556& 1.5059 \\2
&6.9587 &2.9308 & 3 &0.7414 & 1.7274&  3   &0.6606& 1.6083\\ 2.5
&8.5684& 3.4633& 3& 0.7588 &1.562& 2.8444 & 0.7086 &1.4913
\\
  \hline
\end{tabular}\label{tab2}
\end{center}
\end{table*}

\begin{figure}
 \includegraphics[width=0.5\textwidth]{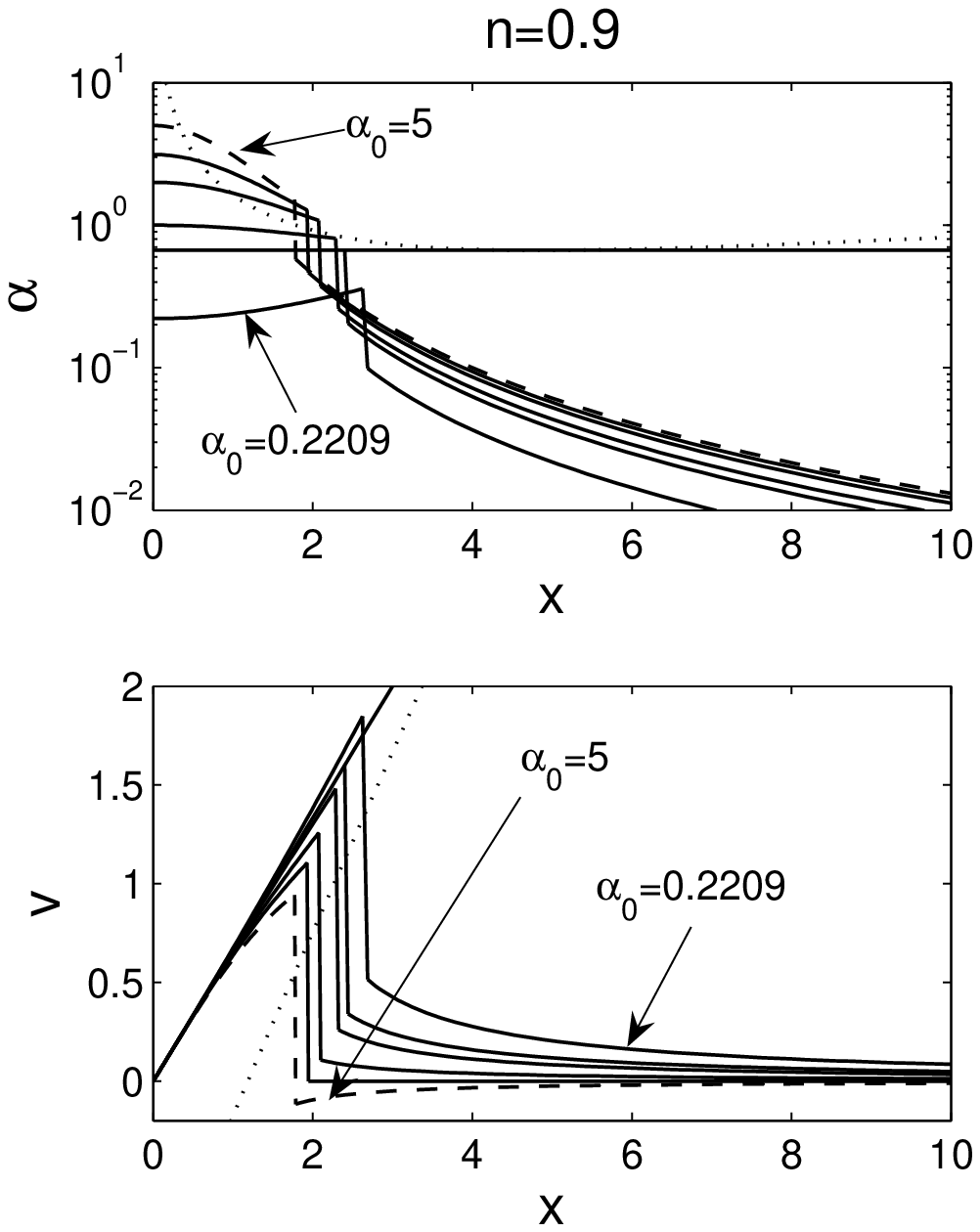}
 \caption{``Champagne flow" solution with a LP type downstream
and the upstream part as a breeze or contraction for $n=0.9$ (i.e.,
$\gamma=1.1$). The top panel shows the density and the bottom panel
shows the velocity. In both panels the dotted curve is the sonic
critical curve. The solutions are integrated from $\alpha_0=0.2209$,
$2/3$, $1$, $2$, $3.13$ and $5$ (from bottom to top in the top panel
and from top to bottom in the bottom panel), and the downstream
shock positions $x_{s{\rm d}}$ are carefully chosen to let the
upstream solution correspond to the asymptotic solutions with $B=0$.
Relevant parameters are summarized in Table \ref{tabb}. For the
solution with $\alpha_0=3.13$, the corresponding upstream is a SPS.
For $\alpha_0>3.13$, the upstream contracts (dashed curve), while
for $\alpha_0<3.13$, the upstream is a breeze. For a breeze
``champagne flow", we require $\alpha_0\geqslant3.13$ (solid
curves).}
 \label{Breeze}
\end{figure}

In the isothermal case of \citet{Shu}, the solution in which the
outer part is a static SIS represents a limiting solution defining
the maximum value of $\alpha_0$. In the polytropic cases with
$n=0.9$, we can also identify such a limit by requiring $B=0$ for
the upstream asymptotic solutions, such that for a preset
$\alpha_0$ the downstream shock position $x_{s{\rm d}}$ is
uniquely determined. A family of such ``champagne flow" solutions
with asymptotic upstream breezes or contractions is shown by solid
curves in Figure \ref{Breeze} with parameters summarized in Table
\ref{tabb}. With a gradual increase of $\alpha_0$, the upstream
side varies from outward breezes, to a SPS, and to an inward
contraction; this trend leads to a maximum $\alpha_0$ if we define
an outward breeze for classical ``champagne flows". Here, the
critical value $\alpha_0=3.13$ for an upstream SPS corresponds to
the upstream SIS limit in \citet{Shu} (i.e., $\alpha_0=7.9$ in the
isothermal case). Naturally, this critical value depends on the
choice of $n$. For $\alpha_0>3.13$, an upstream asymptotic
contraction or an ISECE solution appears (dashed curve in Fig.
\ref{Breeze}). Thus, isothermal champagne flows
form a special family with $B=0$ and $n=1$, referred
to as breeze champagne flows. With $2/3<n<1$ or $2<l<3$, there are
many more physically possible champagne flows, for which the
upstream side corresponds to asymptotic outflows at large $x$.
The physical meaning of this special solution whose upstream side is
the outer part of a static SPS is clear: the outer envelope of gas
is initially in a hydrostatic equilibrium, and the expanding shock
created by the UV photoionization travels into the static envelope;
on the downstream side of this expanding shock, the gas is heated to
high temperatures. As the static SPS relies on the scaling parameter
$n$ or the polytropic index $\gamma$, we expect one single solution
for a preset $n$. From Table \ref{tabb} we see clearly that with the
increase of $\alpha_0$, the downstream shock position to obtain a
breeze in the upstream, or $x_{\rm min2}$ decreases. Meanwhile,
numerical explorations suggest that $x_{\rm min1}$ increases with
increasing $\alpha_0$. Hence we expect for a sufficiently large
$\alpha_0$, $x_{\rm min2}$ would become less than $x_{\rm min1}$ to
forbid ISECE solutions.

\begin{table*}
\begin{center}
  \caption{Parameters of Global Polytropic ``Champagne Flow" Solutions
  with $n=0.9$ and Upstream Breeze or Contraction ($B=0$).
  }
  \begin{tabular}{|l|l|l|l|l|l|l|l|l|}
  \hline
  \hline
  $\alpha_0$ & $A$ & $B$ & $x_{s{\rm d}}$ & $\alpha_{\rm d}$ & $v_{\rm d}$
  & $x_{s{\rm u}}$ & $\alpha_{\rm u}$ & $v_{\rm u}$\\
   \hline
0.2209&0.7550&0&2.6250&0.3571&1.8473&2.6873&0.0990&0.5153\\
2/3   &1.3240&0&2.3986&0.6667&1.5991&2.4419&0.2045&0.3407\\
1     &1.5357&0&2.2869&0.8068&1.4805&2.3234&0.2586&0.2599\\
2     &1.8643&0&2.0733&1.0783&1.2563&2.0996&0.3734&0.1069\\
3.13  &2.0420&0&1.9264&1.2773&1.1036&1.9475&0.4643&0\\
5     &2.1924&0&1.7706&1.5091&0.9429&1.7873&0.5746&$-0.1163$\\
  \hline
\end{tabular}\label{tabb}
\end{center}
\end{table*}

\section[]{Similarity Polytropic ``Champagne Flows" with Central Voids}


We now establish and analyze a new class of ``champagne flow"
solutions with voids surrounding the centre. We extend solutions
from $x=0$ to the dimensionless self-similar expanding boundary
$x^{\ast}$ of a void, inside of which there is no mass, i.e.,
$m^*=m(x^{\ast})=0$.
%
%
In our notations, superscript $^{\ast}$ attached to variables
indicates variables on the void boundary $x^{\ast}$. By expression
(\ref{equ7}), we have $v^{\ast}=nx^{\ast}$.
The void boundary conditions are
\begin{equation}
\alpha=\alpha^*\ ,\qquad v=nx^*\ ,\qquad\hbox{at}\qquad x=x^*\ .
\label{equ20}
\end{equation}
A Taylor series expansion to the first order around the void
boundary $x=x^*$ of ODEs (\ref{equ9}) and (\ref{equ10}) yields
\begin{eqnarray}
&&v(x)=nx^{\ast}+2(1-n)(x-x^{\ast})+\cdots\ ,\qquad
\label{equ23}\\
&&\alpha(x)=\alpha^*+\frac{n(1-n)}{\gamma} (\alpha^*) ^n
x^{\ast}(x-x^{\ast})+\cdots\ . \label{equ24}
\end{eqnarray}
Series expansions (\ref{equ23}) and (\ref{equ24}) are
conspicuously different from series expansions (\ref{equ13}) and
(\ref{equ14}). The proportional coefficient of the asymptotic
velocity is no longer $2/3$ but depends on the scaling index $n$.
By translating the origin, the dynamic flow behaves differently. For
$n<2$, solutions given by expression (\ref{equ23}) are locally
above the line $nx-v=0$, indicating a positive enclosed mass.
Numerical integrations reveal that the solutions are always above
the line $nx-v=0$ thereafter.
Here, we model a gas flow with a central void in the presence of
self-gravity and thermal pressure, directly relevant to galactic
subsystems such as H \Rmnum{2} regions.

\begin{figure}
 \includegraphics[width=0.5\textwidth]{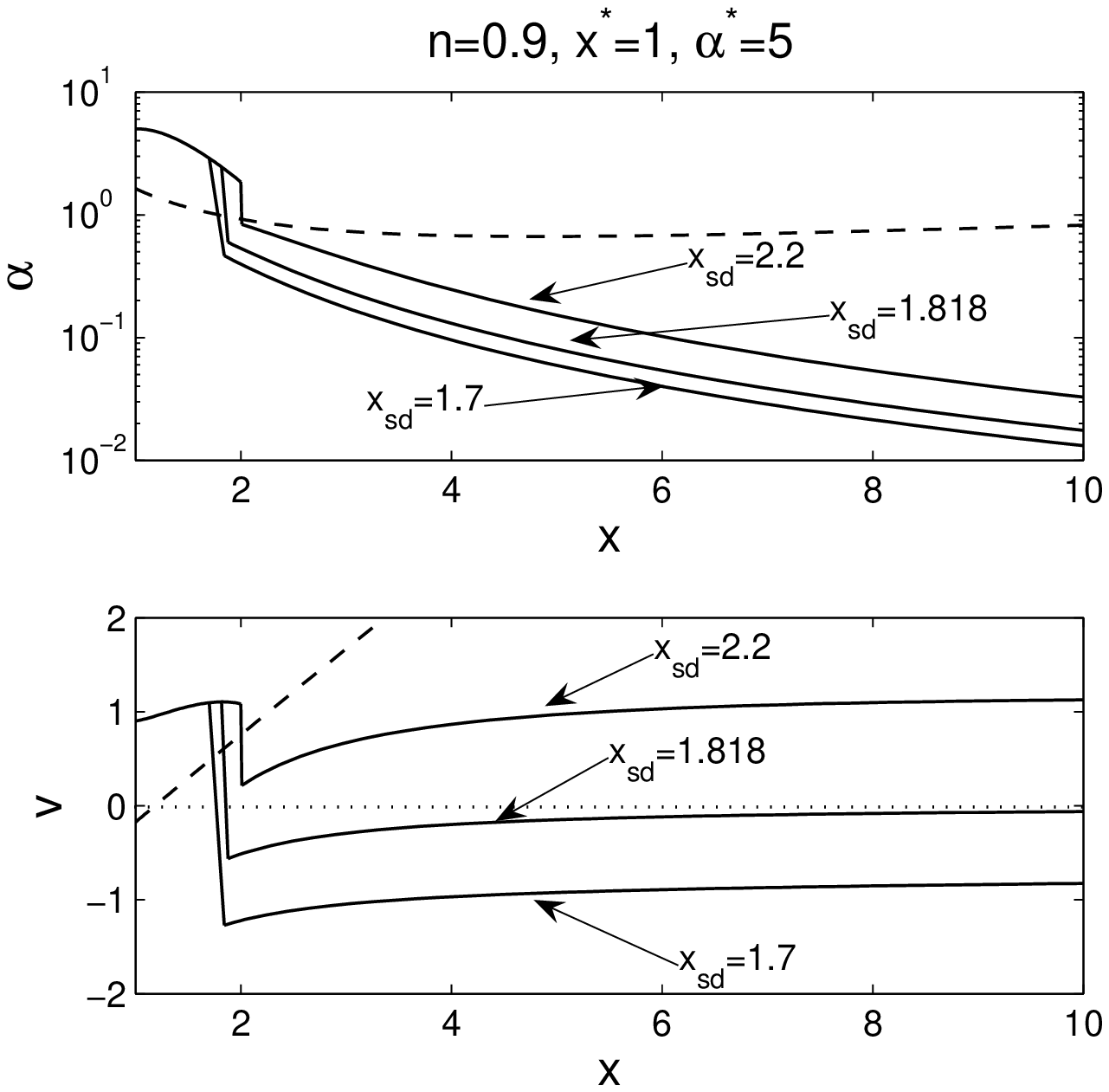}
 \caption{The reduced mass density $\alpha(x)$ (top) and the reduced
 radial flow velocity $v(x)$ (bottom)  for semi-complete ``champagne
 flow" solutions with a central void inside $x^{\ast}=1$ in the case of
 $n=0.9$ and density on the void boundary $\alpha^{\ast}=5$. In both
 panels, the dashed curve stands for the sonic critical curve. In the
 bottom panel the dotted line stands for the line $v=0$. The solid curve
 on the upper left of the sonic critical curve is the downstream void
 solution and the solid curves on the lower right of the sonic critical
 curve are the corresponding upstream solutions with the downstream shock
 position $x_{s{\rm d}}=$ 1.7 (inflow), 1.818 (contraction)
 and 2 (outflow).
 In this case, $x_{\rm min2}\approx 1.818$ and $x_{\rm min1}=1.408$. 
 Numerical data about these solutions are tabulated in Table
 \ref{tab3}.}
 \label{Fig4}
\end{figure}

\begin{figure}
 \includegraphics[width=0.5\textwidth]{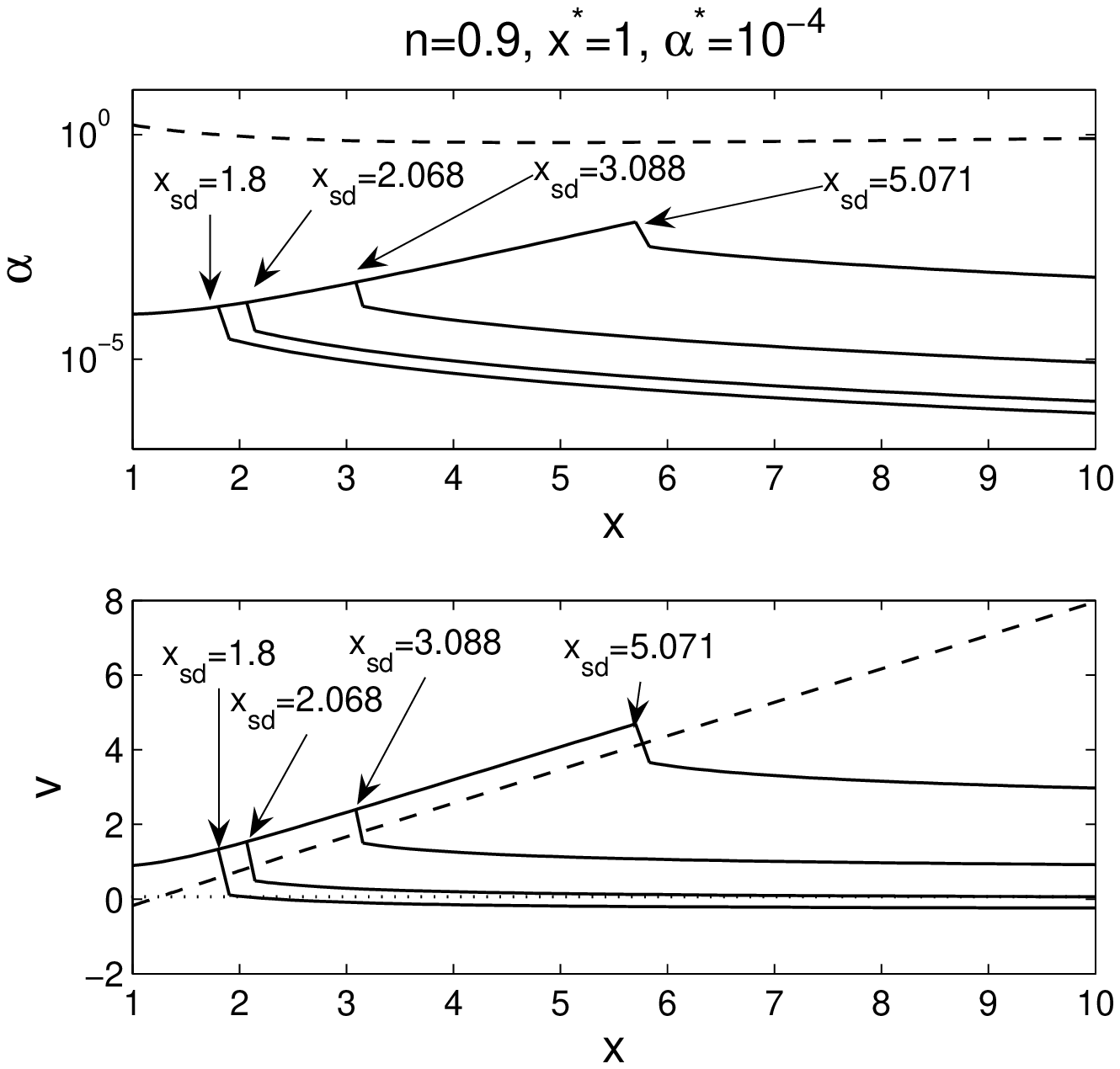}
 \caption{The reduced mass density $\alpha(x)$ (top) and the reduced
 radial velocity $v(x)$ (bottom) for semi-complete ``champagne flow"
 solutions with a central void inside $x^{\ast}=1$ in cases with $n=0.9$
 (thus $\gamma=1.1$) and a density on the void boundary $\alpha^{\ast}=
 10^{-4}$. In both panels, the dashed curve represents the sonic critical
 curve. In the bottom panel, the dotted line stands for the line $v=0$.
 The solid curve on the left side is the downstream void solution and the
 solid curves on the right side are the corresponding upstream outflow
 solutions with the downstream shock position $x_{s{\rm d}}=1.8$ (inflow),
 2.068 (breeze), 3.088 (outflow) and 5.701 (outflow).
 Here, $x_{s{\rm d}}=2.068$ is the limit to ensure an asymptotic outflow,
 i.e., $x_{\rm min2}\approx 2.068$. In this case, $x_{\rm min1}=1.261$.
 Numerical data for these solutions are tabulated in Table \ref{tab3}.}
 \label{Fig5}
\end{figure}

One can readily obtain the downstream portion of ``champagne flow"
solutions by numerically integrating coupled nonlinear ODEs
(\ref{equ9}) and (\ref{equ10}) from the void boundary $x^{\ast}$
with asymptotic expansions (\ref{equ23}) and (\ref{equ24}).
%
%
%
All such void solutions encounter the sonic critical curve, and
the lower the value of $\alpha^{\ast}$ is, the later (or at larger
$x$) the void solution encounters the sonic critical curve. In
order to match with asymptotic solutions of finite density and
velocity at large $x$, void solutions must either cross the
critical curve smoothly or connect to another branch of solutions
via shocks. To model ``champagne flows", we need to construct
shocks to obtain global solutions. Global void solutions that
cross the sonic critical curve smoothly will be discussed in Lou
\& Hu (2008, in preparation).
%
%
%
We now present the solutions for large and small $\alpha^{\ast}$,
respectively. A family of semi-complete ``champagne flow"
solutions with $n=0.9$, void boundary $x^*=1$ and $\alpha^*=5$ is
constructed by varying the self-similar shock position as shown in
Figure \ref{Fig4}. Complementarily, another family of solutions
with $\alpha^*=10^{-4}$ is also constructed and shown in Figure
\ref{Fig5}. Both outflow, inflow and breeze or contraction as the
upstream side are presented. Relevant parameters of these void
``champagne flow" solutions are summarized in Table \ref{tab3}.

\begin{table*}
\begin{center}
  \caption{Polytropic ``Champagne Flow" Solutions with a Central Void inside $x^*=1$ in
  the Case of $n=0.9$}
  \begin{tabular}{|l|l|l|l|l|l|l|l|l|}
  \hline
  \hline
  $\alpha^*$ & $A$ & $B$ & $x_{s{\rm d}}$ & $\alpha_{\rm d}$ & $v_{\rm d}$
  & $x_{s{\rm u}}$ & $\alpha_{\rm u}$ & $v_{\rm u}$\\
   \hline
$10^{-4}$ &0.000106&  $-0.3813$&  1.8&$1.4734\times10^{-4}$& 1.3322
& 1.9058 & $2.7895\times10^{-5}$ & 0.1058\\$10^{-4}$ &0.000191&  0&
2.068 &
$1.8377\times10^{-4}$& 1.5386 & 2.1442 & $4.2673\times10^{-5}$ &  0.4893\\
$10^{-4}$& 0.0014&
1.0833 & 3.088 & 0.000517& 2.398& 3.152&   0.00015& 1.501\\
$10^{-4}$&
0.0991 & 3.5315& 5.701& 0.0114& 4.701& 5.839&   0.0032 & 3.669\\
5& 2.2512& $-1.0512$ & 1.7& 2.8906& 1.0956& 1.8424& 0.4646& $-1.2708$\\
5& 2.9388& 0&
1.818& 2.4535& 1.1047& 1.8777& 0.5978& $-0.5632$\\
5&5.2894&1.7273&2&1.8447&1.0816&2.0101&0.8364&0.2166\\
  \hline
\end{tabular}\label{tab3}
\end{center}
\end{table*}

Similar to no void cases, $\alpha$ decreases with increasing $x$
for large $\alpha^{\ast}$, so the density on the void boundary is
a local maximum (see Fig. \ref{Fig4}); while for small
$\alpha^{\ast}$, $\alpha$ increases with increasing $x$, so the
density maximum is at the downstream side of the shock. The latter
corresponds to a shell-like structure in self-similar expansion
(see Figure \ref{Fig5}).
%
%
%

With downstream void solutions and upstream outflow and breeze
solutions connected by shocks, we establish semi-complete polytropic
``champagne flow" solutions with central voids. Similar to LP type
solutions with shocks, there are also one maximum $x_{\rm max}$ and
two minimum limits $x_{\rm min1}$ and $x_{\rm min2}$ imposed on the
downstream shock position $x_{s{\rm d}}$ in order to obtain
``champagne flow" solutions.
Systematic numerical explorations for cases of $n=0.7$, $n=0.8$ and
$n=0.9$ show that in general $x_{\rm min2}>x_{\rm min1}$. For
$x_{\rm min1}<x_{s{\rm d}}<x_{\rm min2}$, a void solution can be
matched with an asymptotic inflow to produce ISECE solutions, while
for $x_{s{\rm d}}>x_{\rm min2}$, a central void solution can be
matched with an asymptotic outflow to produce ``champagne flow"
solutions.
For $x_{s{\rm d}}=x_{\rm min2}$, the upstream corresponds to a
breeze or a contraction with $B=0$.
The analysis here parallels that in the previous section for cases
without central voids; in particular, the parameters $x_{\rm min1}$
and $x_{\rm min2}$ are determined not only by $n$ and
$\alpha^{\ast}$, but also by the expanding void boundary $x^{\ast}$.
Numerical explorations suggest that for a certain $n$, with the
increase of $\alpha^{\ast}$, $x_{\rm min1}$ increases and $x_{\rm
min2}$ decreases. Hence for a sufficiently large $\alpha^{\ast}$,
we expect $x_{\rm min2}<x_{\rm min1}$ for which ISECE solutions
are not allowed. This is consistent with polytropic cases without
central voids.

\section[]{Analysis and Discussion}

\subsection[]{Comparison with Numerical Simulations}

To adapt our self-similar solutions for modelling an astrophysical
cloud system, we need to first specify the parameter $k$ related to
the sound speed squared. By varying $k$, one can model clouds of
different scales using a single self-similar solution. Parameter $k$
is determined by the thermodynamic parameters, including thermal
pressure $p$, mass density $\rho$ and temperature $T$. A useful
relation for a conventional polytropic gas derived from
transformation (\ref{equ6}) is
\begin{equation}
k=\frac{p}{\rho^{\gamma}(4\pi G)^{\gamma-1}}=\frac{k_B
T}{\mu\rho^{\gamma-1}(4\pi G)^{\gamma-1}}\ ,\label{equk}
\end{equation}
where $\mu$ is the mean molecular mass of gas particles. For a
typical value provided by the classification of \citet{Habing},
the UC H \Rmnum{2} regions have an electron number density
$n_e>3000$ cm$^{-3}$ (corresponding to a mass density
$\rho>5\times10^{-21}$ g cm$^{-3}$ for a fully ionized hydrogen
gas), and the Compact H \Rmnum{2} regions have $1000<n_e<3000$
cm$^{-3}$ (corresponding to
$1.7\times10^{-21}<\rho<5\times10^{-21}$ g cm$^{-3}$ for mass
density $\rho$).
Typically, the temperature of H \Rmnum{2} regions is of order of
$\sim10^4$ K.
For a fully ionized hydrogen gas, we assume that $\mu=m_p/2$ where
$m_p$ is the proton mass. The value of the polytropic index
$\gamma$ does influence very much the resulting $k$ and thus $k$
should be evaluated specifically. For nearly isothermal cases,
with relation (\ref{equk}), we estimate $k$ for UC and Compact H
\Rmnum{2} regions to be $k\sim 10^{11}\sim10^{12}$ cgs unit. One
should be aware that $\kappa$ and thus $k$ vary with the gas
temperature and density in a cloud. Here, we presume a constant
$k$ to convert self-similar variables to real space variables as a
first approximation.

We now compare our self-similar solutions of quasi-spherical
symmetry with previous numerical simulations.
\citet{TT1} performed a numerical study for a similar scenario as
ours, i.e. a nascent central massive protostar ionizes and heats
the ambient neutral gas and then leads to ``champagne flows". In
their simulation, the radiative cooling rate is assumed to be low
and thus our polytropic approach may be applicable.
Franco et al. (1990) gave an analytical model for the formation
and expansion of H \Rmnum{2} regions and have their solutions
compared with simulations. We intend to demonstrate that our
self-similar polytropic analysis of the problem gives
qualitatively similar results.

\begin{figure}
 \includegraphics{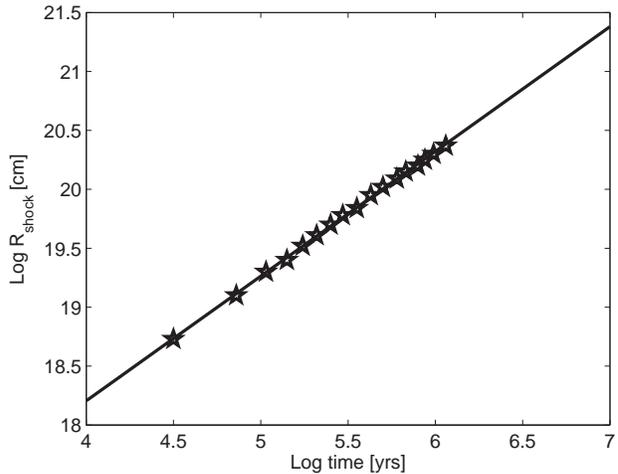}
 \caption{Shock position evolution with time $t$ in a ``champagne flow"
 for our self-similar model (solid line) and the numerical simulation
 of \citet{TT1} (asterisks). For the simulation, the adopted
 parameters are central initial density $\rho_0=2\times10^{-21}$
 g cm$^{-3}$, initial temperature $T_0=3000$ K, and the ionization
 UV flux $F=2\times 10^{51}$ photons s$^{-1}$. For the self-similar
 model, the best fitted parameters are $n=1.0583$ and
 $\log(k^{1/2}x_s)=6.0356$ in cgs unit.
 }
 \label{Fig6}
\end{figure}

In the simulation of \citet{TT1}, computations were carried out
following the progressive ionization of a diffuse gas and
subsequent dynamical evolution, in a globular cluster soon after
the star formation has been initiated. The residual gas is
initially in a hydrostatic equilibrium in the gravitational field
at a uniform temperature and all the ionizing UV radiation comes
from stars at the very centre of the cluster.
We emphasize that only the cases in
which the gas is fully ionized correspond to our conventional
``polytropic champagne flow" model.
In the simulation, the initial mass density at the centre
$\rho_0$, initial temperature $T_0$, and the UV photon flux $F$
from the central star completely determine the evolution in
spherical symmetry. To translate these parameters into our
self-similar form of solutions, we first explore the time
evolution of a ``champagne flow" shock. In our self-similar model,
the shock radius $r_s$ obeys $r_{s}=k^{1/2}x_st^n$.
With such a self-similar formula, we can fit the scaling index $n$
and $k^{1/2}x_s$ to shock positions obtained by the numerical
simulation for case F of \citet{TT1},
and the result comparison is shown in Figure \ref{Fig6} with
relevant parameters in caption. We see an almost perfect fit,
suggesting that the dynamical evolution of ``champagne flows"
approaches a self-similar form. This fitting gives a value of
$n=1.0583$ and $\log(k^{1/2}x_s)=6.0356$ in cgs unit (strictly
speaking, we need a general polytropic gas for $\gamma>1$). The
value of $n$ is fairly close to the isothermal case of $n=1$,
consistent with the model analysis of the numerical simulation
that the gas evolution is nearly isothermal \citep{TT1}.

\begin{figure}
 \includegraphics[width=0.5\textwidth]{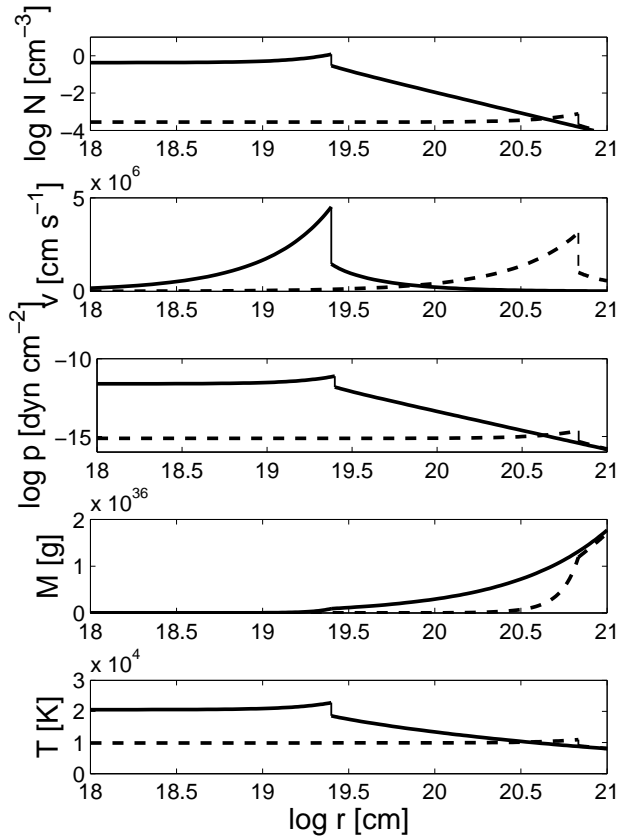}
 \caption{
 Self-similar ``champagne flow" solutions for radius up to
 $10^{21}$ cm ($\sim 300$ pc) at time $t=1.3\times 10^5$ yr (solid curves) and
 $t=5.1\times10^6$ yr (dashed curves). From top to bottom, the panels show number density,
 velocity, pressure, enclosed mass and temperature of the gas, respectively.
 The self-similar shock solution is obtained with
 $n=0.9$, $\gamma=1.1$, $\alpha_0=1\times10^{-5}$ and downstream shock
 position $x_{s{\rm d}}=1.86$. The downstream sound scaling factor $k_{\rm d}$ is
 $3.6\times10^{15}$ cgs unit, and the upstream sound scaling factor $k_{\rm u}$ is
 $3.38\times10^{15}$ cgs unit. The self-similar variables on the downstream side of
 the shock are $(x_{s{\rm d}},\ \alpha_{\rm d},\ v_{\rm d})=(1.86,\ 2.80\times10^{-5},
 \ 1.37)$, and the corresponding upstream variables are $(x_{s{\rm u}},\ \alpha_{\rm u},\
 v_{\rm u})=(1.92,\ 6.88\times10^{-6},\ 0.46)$. At large $x$, the numerical solution
 matches with asymptotic solution (\ref{equ15}) and (\ref{equ16}) with $A=31.942$ and
 $B=1.006$ at large $x$.}\label{Fig7}
\end{figure}

We now generate a global ``champagne flow" solution grossly
comparable to figure 2 of \citet{TT1} by fitting parameters. We
first choose a central reduced density
$\alpha_0\sim1\times10^{-5}$ such that the initial central density
yields the value used in the simulation. We expediently choose
$n=0.9$ (hence the parameter $\gamma=2-n=1.1$) to model the
initial density profile $l=-2/n=-2.22$. Note that parameter $n$
for the ``champagne flow" solution is slightly different from the
value we obtain from the fitting.
With $\rho_0$ and $T_0$ specified in the simulation, we estimate
$k_d=3.6\times 10^{15}$ cgs unit with expression (\ref{equk}).
We still have the freedom to require the shock traveling to
$r_s=2.51\times10^{19}$ cm at $t=1.3\times10^5$ yr, giving
$x_{s{\rm d}}=1.86$. With these parameters, we model ``champagne
flows" in diffuse H \Rmnum{2} regions with radius $r$ up to
$10^{21}$ cm (i.e., $\sim$ 300 pc). The full solution is shown in
Figure \ref{Fig7}. The timescale of $1.3\times10^5$ yr is regarded
as the duration of the formation phase and the initial time for a
``champagne flow", and the timescale $5.1\times10^6$ yr is
regarded as the lifetime of a ``champagne flow".

The orders of magnitude of all variables are consistent with
typical values; e.g., the expansion velocity is several tens km
s$^{-1}$ and the temperature is about $\sim 10^4$ K.
The enclosed mass at $r=10^{21}$ cm ($\sim 300$ pc) is about $850
M_{\odot}$, consistent with the value of $\sim 800 M_{\odot}$
given by the numerical simulation and with the typical value for
diffuse H \Rmnum{2} regions. The enclosed mass does not vary with
time $t$, confirming the cut-off radius chosen at $r=10^{21}$ cm.
As time evolves, the central number density decreases from
$10^{-0.5}$ to $10^{-3.5}$ cm$^{-3}$ and the central thermal
pressure decreases accordingly from $10^{-12}$ to $10^{-15}$ dyne
cm$^{-2}$.

We can also compare variable profiles with the case F of
\citet{TT1}. The velocity profiles are very similar and we clearly
see an expanding shock. As time evolves, the shock strength
becomes weaker. In both numerical simulation and our model
analysis, we observe a density peak on the downstream side of the
shock and a significant temperature gradient on the upstream side,
which cannot be accounted for by previous isothermal solutions.
The upstream density and pressure profiles are also similar;
however, the downstream density, pressure and temperature profiles
(near centre) are somewhat different. In \citet{TT1}, temperature,
pressure and density are initially very uniform behind the
champagne shock but at the end of the calculations show large
inward gradients. In Figure \ref{Fig7}, our model also produces
quasi-uniform temperature, pressure and density behind the shock
at the beginning ($t=1.3\times10^5$ yr), but we do not observe
large inward gradients at the end. This differences are primarily
due to the different physical assumptions adopted in the
simulation and our self-similar solutions. \citet{TT1} treated the
gas dynamics in protoglobular clusters and neglected the gas
self-gravity as the gas mass is only about 0.1 per cent that of
the stars. As shown above, our self-similar solutions neglect the
gravity of the central massive protostar but include the
self-gravity effect. Another important factor that introduces such
difference is that in the simulation both forward champagne shock
and reverse rarefaction wave are taken into account. Our
self-similar model can accommodate forward moving shock, so we
only have the principle outgoing champagne shock.

We calculate the total energy $E_{total}$ defined as the energy of
the gas under consideration in an infinite space. The total energy
at time $t$ can be expressed as
\begin{eqnarray}
\!\!\!
&E_{\rm total}=E_{\rm K}+E_{\rm G}+E_{\rm I}
\qquad\qquad\qquad\qquad\nonumber\\
&=\int_{r_{\rm in}}^{r_{\rm out}}\bigg(\frac{1}{2}\rho
u^2-\frac{GM\rho}{r}+\frac{i}{2}p\bigg)4\pi
r^2dr\qquad\nonumber\\
&=\frac{k^{5/2}t^{5n-4}}{2G}\int_{x_{\rm in}}^{x_{\rm
out}}\bigg[\alpha
v^2x^2-\frac{2}{(3n-2)}\alpha^2x^3(nx-v)\nonumber\\
&+i\alpha^{\gamma}x^2\bigg]dx\ ,
\end{eqnarray}
where $E_K$, $E_G$ and $E_I$ are the kinetic, gravitational and
internal energies of the gas, respectively, and $r_{\rm in}$,
$r_{\rm out}$, $x_{\rm in}$, $x_{\rm out}$ are the inner and outer
boundaries of the gas under consideration, $i$ is the degree of
freedom of the gas particle presumed to be 3. Note that one fixed
$r_{\rm out}$ at different times corresponds to different values
of $x_{\rm out}$. In this solution, $E_{\rm
total}=5.5\times10^{48}$ erg at $t=1.3\times10^5$ yr, and $E_{\rm
total}=8.1\times10^{48}$ erg at $t=5.1\times10^6$ yr, so there is
net energy input. Especially, we see the kinetic energy
$E_K=6.4\times10^{47}$ erg at $t=1.3\times10^5$ yr, indicating a
small fraction of the total energy, and $E_K=3.9\times10^{48}$ erg
at $t=5.1\times10^6$ yr, indicating a fairly large fraction of the
total energy. The increase of kinetic energy shows clearly the
development of a champagne flow. The gravitational energy is of
order $10^{44}$ erg, much less than the kinetic energy throughout
this duration. This confirms that the gas is not bounded and must
have an outflow. We are also able to consider qualitatively the
local energy exchange throughout the self-gravitating gas with
relation (\ref{Econ}). In Figure \ref{Fig7}, $\partial p/\partial
r$ is positive on the downstream side and negative on the upstream
side. For $\gamma=1.1$, the downstream and upstream sides locally
loses and gains energy, respectively.
%
In summary, the profiles on the order of magnitudes, and the time
evolution of our self-similar solution in modelling this case are
grossly consistent with the numerical simulation result of
\citet{TT1}, which lends support to our polytropic self-similar
``champagne flow" solution as a gross description of dynamics of H
\Rmnum{2} regions.

Recent studies further suggest that the inclusion of stellar winds
is also important and even necessary sometimes in understanding
the large-scale dynamics of H \Rmnum{2} regions. \citet{Comeron}
found that a shocked stellar wind in the central region produces
important morphological differences as compared to windless cases.
Moreover, \citet{Comeron} suggested that the spatial scale of an H
\Rmnum{2} region undergoing ``champagne flow" is systematically
larger and the gas flow is generally faster as driven by a central
stellar wind. \citet{Arthur} provided two-dimensional cylindrical
radiative-hydrodynamic simulations of cometary H \Rmnum{2} regions
using champagne flow models,
by taking into account of strong stellar winds from the central
ionizing star. In these simulations, the hydrodynamics and
radiative transfer are coupled through an energy equation whose
source term depends on the photoionization heating and radiative
cooling rates; while with our polytropic approach, complicated
energetic processes are relegated to the choice of $\gamma$.
\citet{Arthur} studied the hydrodynamics of a compact H \Rmnum{2}
region with a radius 0.13 pc and at a time $\sim 200$ yr after the
triggering of UV ionizing photons and powerful stellar winds; a
stellar wind bubble around the centre with a radius up to 0.03 pc
is formed. Inside such a stellar wind bubble, the mass density is
about 3 orders of magnitude lower than that of the surrounding
medium, and the density of the flow does not vary much with radius
in the vicinity of the bubble boundary. Because the central wind
bubble is effectively depleted of mass and the gravity force of
the central massive star may be neglected, given a typical
Bondi-Parker radius of $\sim 10^2$ AU, we thus approximate such a
stellar wind bubble as a central `void' and model it using our
polytropic self-similar void ``champagne flow" model.

In the scenario as outlined by \citet{Arthur}, the central star
has an effective temperature $T_{\rm eff}=3\times10^4$ K, a
stellar wind mass-loss rate $\dot{M}=10^{-6}$ M$_{\odot}$
yr$^{-1}$ and a terminal wind speed $V_{\rm w}=2000$ km s$^{-1}$.
The initial ambient medium has a number density of $n_0=6000$
cm$^{-3}$ and a temperature of $T_0=300$ K. In our self-similar
model for ``champagne flows", the radius of a void boundary is
$r^*=k^{1/2}x^*t^n$. By taking $n=0.8$, we have
$k^{1/2}x^*=1.34\times10^9$ cgs unit to obtain a $r^*=0.03$ pc
central void at a time of $t=200$ yr. The $n$ value depends on the
energetic process, including plasma cooling and radiative heating,
of the flow. We further estimate from relation (\ref{equk}) for a
downstream sound parameter $k_d=2.5\times10^{17}$ cgs unit, and
for a self-similar void boundary $x^{\ast}=2.68$. Another
parameter that needs to be specified is the mass density on the
expanding void boundary, denoted as $\alpha^{\ast}$ here. The
simulation of \citet{Arthur} gives an electron number density on
the void boundary as $n_e^{\ast}=10^4$ cm$^{-3}$ at $t=200$ yrs.
With relation $\rho^{\ast}=\alpha^{\ast}/(4\pi G t^2)$, we
estimate $\alpha^{\ast}=5.5\times10^{-7}$. We emphasize that the
length and time scales in this case are quite different from those
in the previous case of \citet{TT1}. As an important advantage,
this suggests that self-similar models are suitable to give a
unified description for cloud systems on quite different scales.

\begin{figure*}
 \includegraphics[width=\textwidth]{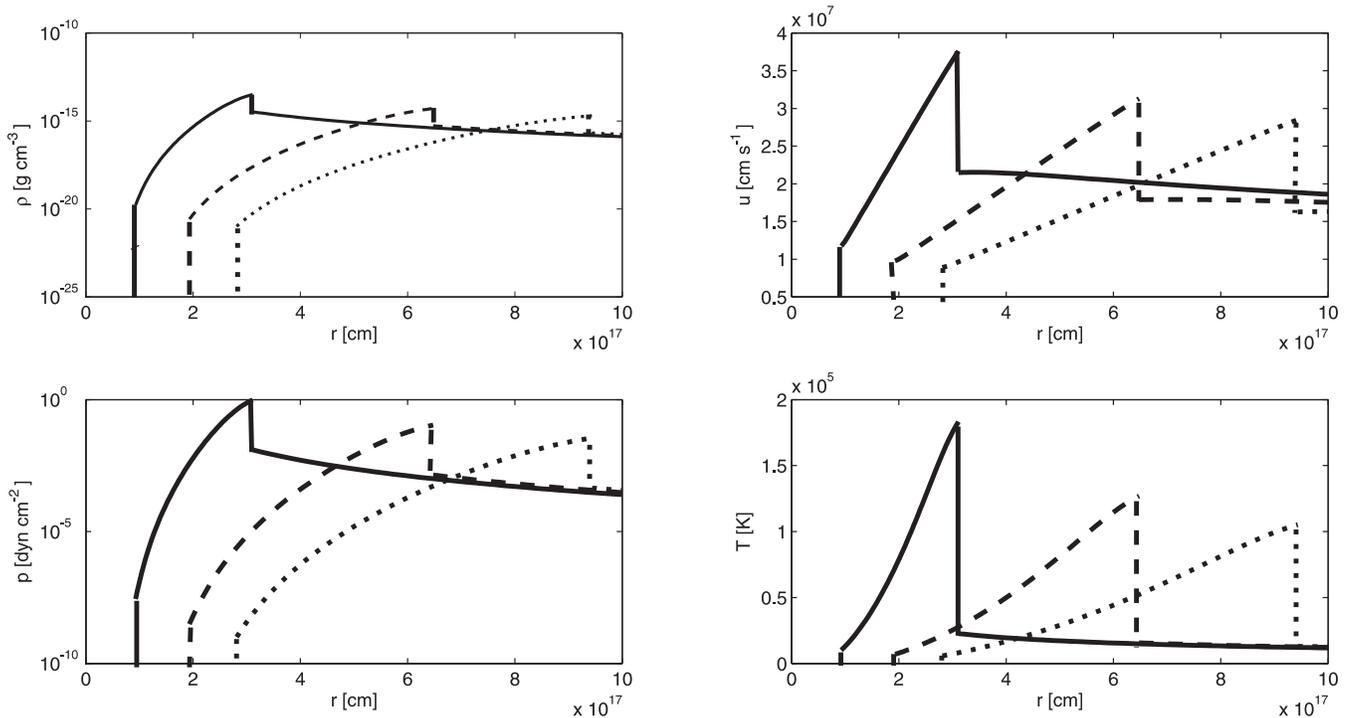}
 \caption{Self-similar ``champagne flow" solution with
 an initial central void radius of $10^{17}$ cm at
 time $200$ yr (solid curve), $500$ yr (dashed curve) and $800$ yr
 (dotted curve). The four panels show mass density $\rho$, flow velocity
 $u$, thermal pressure $p$, and temperature $T$ of the gas,
 respectively. The central void has a radius of $x^{\ast}=2.68$,
 corresponding to $r^{\ast}=0.03,\ 0.06,\ 0.09$ pc with increasing
 time $t$. The self-similar solution is obtained with
 parameters: $n=0.8$, $\gamma=1.2$, $\alpha^{\ast}=5.5\times10^{-7}$
 and downstream shock position $x_{s{\rm d}}=9$. The downstream
 sound parameter $k_{\rm d}$ is $2.5\times10^{17}$ cgs unit,
 and the upstream sound parameter $k_{\rm u}$ is $4.9\times10^{16}$
 cgs unit. The self-similar variables on the downstream
 side of the shock are $(x_{s{\rm d}},\ \alpha_{\rm d},\ v_{\rm d})
 =(9,\ 1.07,\ 6.86)$, and the corresponding upstream variables are
 $(x_{s{\rm u}},\ \alpha_{\rm u},\ v_{\rm u})=(20.4,\ 0.111,\ 8.88)$.
 At large $x$, the solution matches with asymptotic solution
 (\ref{equ15}) and (\ref{equ16}) with $A=135$ and $B=22$ at large $x$.
 The enclosed mass by radius of $10^{18}$ cm is $1.38\times10^6$,
 $1.23\times10^6$ and $1.06\times10^6$ M$_{\odot}$ as time evolves.
 The total energy of gas is $5.86\times10^{53}$, $5.94\times10^{53}$
 and $5.98\times10^{53}$ erg, respectively.
 }
 \label{Fig8}
\end{figure*}

As $n<1$ in this case, we have one more degree of freedom to
specify the shock position. In principle, the shock position is
determined by both the initial density (mass parameter $A$) and
the initial gas motion (velocity parameter $B$). We find that the
lower limit of the downstream shock position $x_{s{\rm d}}$ is
$8.8$, according to condition (\ref{xmin}); thus, the minimum
shock position at $t=200$ yrs is $\sim 3.04\times 10^{17}$ cm
(i.e., about $0.1$ pc). The numerical simulation of \citet{Arthur}
studied the gas dynamics up to a radius of $0.13$ pc. As an
example of illustration, we assume $x_{s{\rm d}}=9$ and show the
resulting solution in Figure \ref{Fig8}. This solution clearly
shows that as time evolves, the void boundary expands, meanwhile
the density and pressure in the vicinity of void boundary decrease
by several orders of magnitude, consistent with the simulation.
However, we see a very high density near the ``champagne flow"
shock on the downstream side, and as time evolves, the density
profile becomes more and more smooth.
For our self-similar ``champagne flow" model with central
expanding voids, the velocity can rise up to several hundred km
s$^{-1}$. In Figure \ref{Fig8}, we also see clearly that the case
is non-isothermal, and on the downstream side of the shock the
temperature is the highest as expected. We note that at $t=800$
yrs, the shock is at $\sim 0.3$ pc, beyond the scale of UC or
compact H \Rmnum{2} regions. In reality, a champagne shock is so
fast that even at a short timescale of $\sim 800$ yrs the shock is
well in the surrounding diffuse interstellar medium (ISM).

Compared with numerical simulations, the advantage of our
semi-analytical self-similar approach is clear. We can generate
self-similar shock solutions to model different H \Rmnum{2}
regions by varying a few parameters. The self-similar processes
and shock solutions of this paper describe the basic hydrodynamics
of polytropic ``champagne flows" and serve as test cases for bench
marking numerical simulations.

\subsection[]{Asymptotic Free-Fall Solutions around a Central Protostar}

\begin{figure}
 \includegraphics[width=0.5\textwidth]{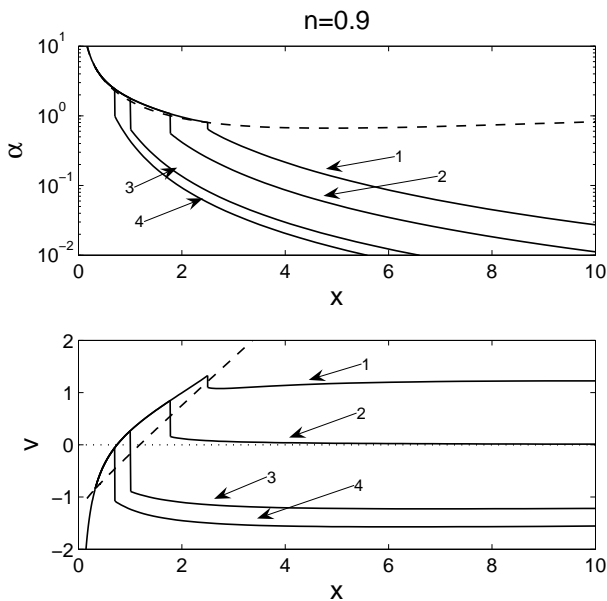}
 \caption{Reduced mass density $\alpha(x)$ (top) and reduced
 radial flow velocity $v(x)$ (bottom) for global solutions in cases
 with $n=0.9$ (thus $\gamma=1.1$) whose downstream side is free-fall
 solution and the upstream side corresponds to either outflow, inflow,
 breeze or contraction. In both panels, the dashed curve represents the
 sonic critical curve; in the bottom panel the dotted line is $v=0$. The
 downstream solution is connected with the upstream solutions with solid
 curves via shocks. The downstream solution is integrated
 from a sonic critical point $(x,\ \alpha,\ v)=(0.3237,\ 5.0050,\ -0.8455)$
 towards $x\rightarrow 0^{+}$ with a central free-fall asymptotic solution
 of $m(0)=0.546$, and outwards to the downstream shock positions. At the
 inner most part the downstream solution corresponds to an inflow, and
 outer part of the downstream side is an outflow. The static point in this
 case is at $x_{static}\sim 0.74$. In both panels, the upstream solutions
 from top to bottom correspond to $x_{s{\rm d}}=2.5$ (labeled 1), 1.7747
 (labeled 2), 1 (labeled 3) and 0.7 (labeled 4). Solution 1 has inner
 inflow and outer outflow on the downstream side and upstream outflow with
 $A=4.368$ and $B=1.76$. The shock parameters are $(x_{s{\rm d}},\
 \alpha_{\rm d},\ v_{\rm d})=(2.5,\ 0.8036,\ 1.3249)$, and $(x_{s{\rm u}},\
 \alpha_{\rm u},\ v_{\rm u})=(2.5003,\ 0.6453,\ 1.0981)$. Solution 2 has
 inner inflow and outer outflow on the downstream side and an upstream breeze
 with $A=1.8635$ and $B=0$ at large $x$. The shock parameters are
 $(x_{s{\rm d}},\ \alpha_{\rm d},\ v_{\rm d})=(1.7747,\ 1.0715,\ 0.8470)$
 and $(x_{s{\rm u}},\ \alpha_{\rm u},\ v_{\rm u})=(1.7796,\ 0.5575,\ 0.1558)$.
 Solution 3 has an inner inflow and outer outflow on the downstream side and upstream
 inflow with $A=0.6930$ and $B=-1.6963$ at large $x$. The shock parameters are
 $(x_{s{\rm d}},\ \alpha_{\rm d},\ v_{\rm d})=(1,\ 1.7883,\ 0.2668)$ and
 $(x_{s{\rm u}},\ \alpha_{\rm u},\ v_{\rm u})=(1.0118,\ 0.6348,\ -0.8942)$.
 Solution 4 has a downstream inflow and upstream inflow, with $A=0.4899$ and
 $B=-2.1564$ at large $x$. The shock parameters are
 $(x_{s{\rm d}},\ \alpha_{\rm d},\ v_{\rm d})=(0.7,\ 2.4355,\ -0.0567)$ and
 $(x_{s{\rm u}},\ \alpha_{\rm u},\ v_{\rm u})=(0.7054,\ 0.9837,\ -1.0785)$.}
 \label{Free1}
\end{figure}

\begin{figure}
 \includegraphics[width=0.5\textwidth]{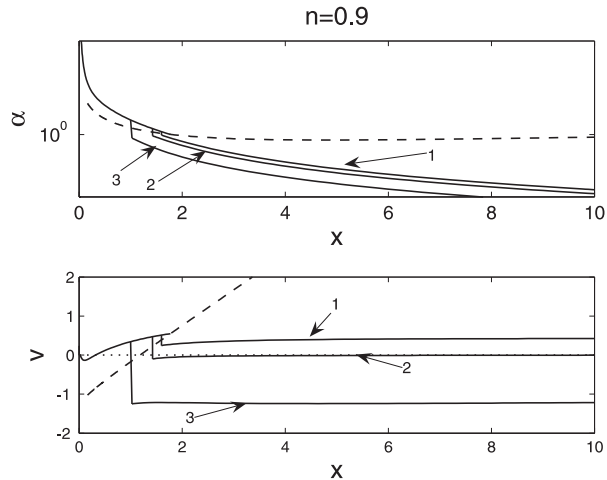}
 \caption{Reduced mass density $\alpha(x)$ (top) and reduced
 velocity $v(x)$ (bottom) for global solutions in cases of $n=0.9$ (thus
 $\gamma=1.1$) whose downstream side is a free fall and upstream
 side corresponds to either outflow, inflow, breeze or contraction. In both
 panels the dashed curve represents the sonic critical curve.
 The downstream solution is connected with upstream solutions of solid
 curves via shocks. The downstream solution
 is integrated from a sonic critical point $(x,\ \alpha,\ v)=
 (1.7727,\ 1.0050,\ 0.5463)$ towards $x\rightarrow 0^{+}$ for a free-fall
 asymptotic solution of $m(0)=4.638$. Most of the downstream side is an
 outflow, while the inner most part is a free fall. In both panels, the upstream
 solutions from top to bottom correspond to $x_{s{\rm d}}=1.6$ (labeled 1),
 1.4269 (labeled 2) and 1 (labeled 3). The entire upstream solution labeled 1
 has an outflow with $A=2.8032$ and $B=0.6058$. Shock
 parameters are $(x_{s{\rm d}},\ \alpha_{\rm d},\ v_{\rm d})=(1.6,\ 1.2326,
 \ 0.5110)$, and $(x_{s{\rm u}},\ \alpha_{\rm u},\ v_{\rm u})=(1.6002,\ 0.9575,\
 0.2443)$. Solution labeled 2 has an upstream contraction with $A=2.115$ and $B=0$
 at large $x$. Shock parameters are $(x_{s{\rm d}},\ \alpha_{\rm d},\
 v_{\rm d})=(1.4269,\ 1.5391,\ 0.4705)$, and $(x_{s{\rm u}},\ \alpha_{\rm u},\
 v_{\rm u})=(1.4290,\ 0.9057,\ -0.0988)$. Solution labeled 3 has an inflow
 for the entire upstream portion with $A=1.0063$ and $B=-1.6727$. Shock
 parameters are
 $(x_{s{\rm d}},\ \alpha_{\rm d},\ v_{\rm d})=(1,\ 2.9346,\ 0.3423)$, and
 $(x_{s{\rm u}},\ \alpha_{\rm u},\ v_{\rm u})=(1.0271,\ 0.7726,\ -1.2514)$.}
 \label{Free2}
\end{figure}

So far we have constructed ``champagne flow" solutions with LP
type asymptotic solutions on the downstream side as $x\rightarrow
0^+$, because such asymptotic solution satisfies boundary
condition (\ref{equ12}). Complementarily, free-fall asymptotic
solutions (\ref{freefall1}) and (\ref{freefall2}) at $x\rightarrow
0^+$ represent gas infall and collapse during the protostar
formation phase; and the surrounding gas and the infall momentum
associated with the star formation process may be sustained for a
while during the evolution after the onset of stellar nuclear
burning and UV photoionization of the surrounding gas.
\citet{Cochran} have investigated consequences of the birth of a
massive star within a dense cloud with a free-fall density
profile, and found that the radiation pressure from the star
sweeps up grains from the infalling gas to form a dust shell which
bounds the H \Rmnum{2} region. Here, we utilize such free-fall
solutions as the downstream side and construct global solutions
with shocks to model possible dynamic evolutions of H \Rmnum{2}
regions surrounding a nascent protostar in nuclear burning. We
present such solutions in Figures \ref{Free1} and \ref{Free2}
where parameter $m(0)$ for free-fall solutions is different. In
dimensionless form, $m(0)$ stands for a central mass point, and
with dimensions in self-similar transformation (\ref{equ6}),
$M(0,t)\propto t^{3n-2}m(0)$; therefore $m(0)$ scales as the
central mass accretion rate. For $m(0)=0.546$ (Figure
\ref{Free1}), the free-fall solution crosses the sonic critical
curve smoothly at $x=0.3237$, and can also be connected to the
upstream solutions via shocks at various locations. The free-fall
solution crosses the line $v=0$ at $x_{\rm stg}=0.74$. This
stagnation radius expands with time in a self-similar manner;
inside the stagnation radius the gas falls inwards, while outside
the stagnation radius the gas expands outwards. Therefore if
$x_{s{\rm d}}<x_{\rm stg}$, the entire global solution corresponds
to an inflow (solution 4 of Figure \ref{Free1}). This situation
describes an accretion shock during a protostar formation phase.
If $x_{s{\rm d}}>x_{\rm stg}$, the outer part of the downstream
side is an outflow. This describes the scenario that the shock
sweeps up the gas and turns the gas from infall to expansion on
the downstream side near the shock front. Similar to the situation
with downstream LP type solutions, there exists one specific
$x_{s{\rm d}}$, from which the upstream solution is a breeze with
$B=0$. In this case, $x_{s{\rm d}}=1.7747$ gives an upstream
breeze (solution 2 of Figure \ref{Free1}). Thus for $x_{s{\rm
d}}<1.7747$, the upstream side corresponds to an asymptotic inflow
far from the centre (solution 3 of Figure \ref{Free1}) and for
$x_{s{\rm d}}>1.7747$, the upstream side corresponds to an
asymptotic outflow (wind) far from the centre (solution 1 of
Figure \ref{Free1}). Another example shown in Figure \ref{Free2}
has $m(0)=4.638$. This free-fall solution does not cross the sonic
critical curve smoothly and can be connected with upstream
solutions via shocks. In an analogous manner, we show the
possibility to obtain an outflow (solution 1 of Figure
\ref{Free2}), breeze (solution 2 of Figure \ref{Free2}) and inflow
(solution 3 of Figure \ref{Free2}) for the upstream side.
Solutions 1 and 2 of Figure \ref{Free1} and solution 1 of Figure
\ref{Free2} have asymptotic outflow or breeze on the upstream side
and in the outer part of the downstream side, which is very
similar to champagne flow solutions obtained with a downstream LP
type, with different behaviours in central regions. With a
free-fall asymptotic solution, the gravity of the central massive
star is not neglected and the gas immediately surrounding the
massive protostar still undergoes infall when the outer envelope
starts to expand. Hence, solutions with free-fall centre are
plausibly suitable to describe the early stage of ``champagne
flows". In general, with central free falls on the downstream
side, we can also obtain asymptotic outflow, inflow, breeze and
contraction for the upstream side, by varying the downstream shock
position $x_{s{\rm d}}$ in a proper range.

Central free-fall solutions describe the core collapse phase in
star formation. For this purpose, \citet{WangLou08} explored such
solutions with free-fall inner core and an inflow or outflow in
the outer envelope, for general polytropic cases in their Figure
2. The inner and outer portions are connected by
magnetohydrodynamic (MHD) shocks. Such shocks are interpreted as
accretion shocks, typically found in a star formation process, or
around accreting black holes. Here we specifically emphasize that
such shocks may also arise by the UV photoionization of ambient
medium surrounding a nascent protostar.
Under certain situations, the UV flux from the burning star might
not be intense and rapid enough to turn the surrounding gas from
infall to expansion by ionization and heating. Meanwhile the
ionization front (IF) creates a weak shock traveling outwards, and
the upstream side may have an outward velocity. Another
possibility is that the gravity of the central star is so large
that the gas immediately surrounding the star keeps falling
towards the protostar, but the outer part of the downstream side
and the corresponding upstream side expand. In summary, with
different initial conditions of gas and different physical
conditions of a burning protostar, the radiative influence of the
nascent protostar on the dynamic evolution of the surrounding gas
may give rise to various self-similar solutions, including the
classical champagne flow solutions, the ISECE solutions and the
inner free-fall with outer inflow/outflows or contraction/breeze
solutions. The solutions constructed in this paper (classical
champagne flows) is suitable for situations that the gas is
initially static and the protostar ionizes the entire gas
immediately, and then the gas begins to expand in a ``champagne
phase" with an outgoing shock.

\section[]{Conclusions}

We present newly established self-similar polytropic shock solutions
with and without central voids to model ``champagne flows" in H
\Rmnum{2} regions featuring various asymptotic dynamic behaviours.
As a substantial generalization of the isothermal model of Tsai \&
Hsu (1995) and \citet{Shu}, we found similarities and differences in
self-similar polytropic processes.
%
Our general polytropic ``champagne flow" model allows a much
larger freedom to choose the polytropic index $\gamma\geq 1$ for
$2/3<n<2$; for a conventional polytropic gas as a subclass of
examples, the power-law index $l$ of the initial mass density
profile $\rho\propto r^{-l}$ is linked to $\gamma$ by
$l=2/n=2/(2-\gamma)$. Together, our model is adaptable to a wide
range of initial mass density profile with $1<l<3$ for
H \Rmnum{2} regions. For conventional polytropic cases of
$1<\gamma<4/3$ (i.e., $2/3<n<1$ and $2<l<3$), we have more
freedom for convergent initial conditions. In this fashion, our
conventional polytropic shock flow solutions are determined not
only by the initial mass density profile (i.e., mass parameter
$A$), but also by the motion at the very early stage (i.e.,
velocity parameter $B$). The dimensionless shock positions or the
dimensional shock speed and strength are determined by the initial
conditions related to $A$ and $B$ and the central density at
$\alpha_0$.
Our self-similar shock flow solutions give a plausible description
for the ``champagne flow" phase for the dynamics of H \Rmnum{2}
regions. We conclude that general polytropic ``champagne flows"
with the initial density power-law index $1<l<3$ may evolve in a
self-similar manner.

We have established novel ``champagne flow" shock solutions with
an expanding void surrounding the centre to model a certain cloud
core whose inner part has fallen into a nascent protostar.
We observe that the evolution of central void boundary plays an
important role in determining the asymptotic solution to approach
and the general behaviour of solutions as well. With even one more
free parameter, the ``champagne flow" shock solutions with central
voids can model the dynamics of H \Rmnum{2} regions more
realistically, including the effect of central stellar wind
bubbles.

We have further explored possibilities of asymptotic inflows or
contractions far from the cloud centre. In addition, we also
establish global shock solutions with the asymptotic free-fall
solution approaching the centre. In general, by varying
dimensionless shock position, we connect the downstream side, with
either LP type solutions, EdS solutions or free-fall solutions, to
upstream solutions which eventually merge into asymptotic outflow,
breeze, contraction and inflow. Within the theoretical framework
of the self-similar polytropic fluid, global shock solutions with
different behaviours correspond to different forms of hydrodynamic
evolution of H \Rmnum{2} regions after the nascence of a central
massive protostar. Apparently, even within the framework of
self-similarity, dynamic evolution of polytropic H \Rmnum{2}
regions depends on the initial and boundary conditions of
molecular clouds. Numerical simulations are needed to probe and
connect various self-similar evolution phases.

\section*{Acknowledgments}

This research has been supported in part
by the Tsinghua Centre for Astrophysics (THCA),
by the NSFC grants 10373009 and 10533020 at Tsinghua University,
and by the SRFDP 20050003088 and the Yangtze Endowment from the
Ministry of Education at Tsinghua University.

\appendix

\section[]{\ \ An Invariant Form }

Transformation (\ref{equ6})
leads to an initial density profile $\rho\propto r^{-2/n}$ as the
outer part of a SPS.
To model an initial mass density profile other than a SIS (e.g.,
Galli et al. 1999) in the self-similarity framework, Shu et al.
(2002) developed an invariant self-similar transformation for an
isothermal gas by neglecting the self-gravity. We extend below
this invariant transformation to a conventional polytropic gas.

The independent similarity variable is
$x=r/(k^{1/2}t^n$) and the reduced radial velocity $v(x)$ still
reads as $v=u/(k^{1/2}t^{n-1})$.
But the mass density $\rho$ is generalized to
\begin{equation}
\rho(r,t)=\frac{D}{r^l}R(x)\ ,\label{M0}
\end{equation}
where $R(x)$ is a new reduced density dependent on $x$ only, $l$ is
a scaling index and $D$ is a constant parameter to be determined.
With the conventional polytropic relation $p=\kappa\rho^{\gamma}$
for constant $\kappa$ and $\gamma$,
the thermal pressure $p$ is
\begin{equation}
p(r,t)=\kappa\frac{D^{\gamma}}{r^{\gamma l}}R^{\gamma}\
.\label{Poly}
\end{equation}
We now determine the relation between $k$ and $\kappa$ from the
radial momentum equation without self-gravity
\begin{equation}
\frac{\partial u}{\partial t}+u\frac{\partial u}{\partial
r}=-\frac{1}{\rho}\frac{\partial p}{\partial r}\ .\label{M1}
\end{equation}
The left-hand side (LHS) of equation (\ref{M1}) relates to the
sound scaling parameter $k$ and the right-hand side (RHS) relates
to the specific entropy coefficient $\kappa$. We substitute
self-similar transformation (\ref{M0}) and (\ref{Poly}) into both
sides of the radial momentum equation (\ref{M1}) to obtain
\begin{equation}\label{leftM}
\frac{\partial u}{\partial t}+u\frac{\partial u}{\partial
r}=k^{1/2}t^{n-2}[(n-1)v+(v-nx)v']\ ,
\end{equation}
\begin{eqnarray}\label{rightM}
-\frac{1}{\rho}\frac{\partial p}{\partial r}=-\kappa
D^{\gamma-1}\gamma
\big(k^{-1/2}t^{-n}\big)^{(\gamma-1)l+1}x^{-(\gamma-1)l}\nonumber\\
\qquad\times\bigg(-\frac{l}{x}R^{\gamma-1}+R'R^{\gamma-2}\bigg)\ ,
\end{eqnarray}
where the superscript `${\prime}$' over $v(x)$ and $R(x)$
indicates the first derivative in $x$. By expressions
(\ref{leftM}) and (\ref{rightM}) and in order to remove the
explicit $t$ dependence in the self-similar form of equation
(\ref{M1}), we require
\begin{equation}
-n[(\gamma-1)l+1]=n-2\ .\label{M2}
\end{equation}
Together with $n+\gamma=2$ for a conventional polytropic gas, we
obtain $l=2/n$ for $n\neq 1$. For $n=1$, there is no constraint on
$l$.
For a conventional polytropic gas, the possible initial mass
density profile described by this invariant self-similar
transformation is not arbitrary, but is preset by the scaling
index $n$ or by the polytropic index $\gamma$. Here $D$ provides a
measure for the magnitude of density profile and $\kappa$ is the
constant in the polytropic equation of state; both parameters are
determined physically. For simplicity of the formulation, we may
choose the sound parameter $k$ so that
\begin{equation}
\kappa=k^{1/n}D^{n-1}\ .
\end{equation}
This relation is verified by a dimensional analysis. As a result,
these parameters disappear in the following ODEs,
\begin{eqnarray}
\big[(nx-v)^2-\gamma x^{2-l} R^{\gamma-1}\big]R'
\qquad\qquad\qquad\qquad\qquad\nonumber\\
\qquad =\frac{Rv}{x}[(l-2)(v-nx)+(n-1)x] -\gamma l
x^{1-l}R^{\gamma}\ ,\label{M3}
\end{eqnarray}
\begin{eqnarray}
\big[(nx-v)^2-\gamma x^{2-l} R^{\gamma-1}\big]v'
\qquad\qquad\qquad\qquad\qquad\nonumber\\
\ \ \quad =2\gamma x^{1-l} R^{\gamma-1}(v-x)+(n-1)v(nx-v)\
.\label{M4}
\end{eqnarray}
For $n\neq 1$ and $l=2/n$, coupled equations (\ref{M3}) and
(\ref{M4}) describe non-isothermal self-similar flows of a
conventional polytropic gas. For $n=1$ and $\gamma=1$, the value
of $l$ is arbitrary; for $l=2$, these two coupled nonlinear ODEs
reduce to the isothermal formulation of Shu et al. (2002) with two
decoupled ODEs for $l=2$. For $n=1$ and $\gamma=1$ with $l\neq 2$,
we need to use the two decoupled ODEs $(32)$ and $(33)$ in Section
4 of Shu et al. (2002).
In contrast to the isothermal case, the nonlinear ODEs here remain
coupled for $\gamma\neq1$. In summary, with our self-similar
transformation,
hydrodynamics for a spherical gas with ignorable gravity can be
expressed by a set of two coupled nonlinear ODEs (\ref{M3}) and
(\ref{M4}).

In cases with $\gamma\neq1$ and for $x\rightarrow +\infty$, the
asymptotic boundary conditions are $v\rightarrow 0$ and
$R\rightarrow 1$ for $n<1$. The asymptotic solution at large $x$
yields
\begin{eqnarray}
v=Hx^{1-1/n}+\frac{2\gamma}{n}x^{1-2/n}+\cdot\cdot\cdot\ ,\qquad\nonumber\\
R=1+3H\bigg(\frac{1}{n}-1\bigg)x^{-1/n}+\cdot\cdot\cdot\ ,\qquad
x\rightarrow +\infty\ ,\label{InLargeX}
\end{eqnarray}
where $H$ is an arbitrary constant but should be chosen for a
positive mass density everywhere. The condition $l=2/n$ has
already been taken into account in this solution. This form of
asymptotic solution at large $x$ differs from equation (34) of Shu
et al. (2002), because for non-isothermal cases the leading terms
in nonlinear ODEs (\ref{M3}) and (\ref{M4}) are different;
however, the second term on the RHS of the expression of $v$,
reduces to the $v$ expression in equation (34) of Shu et al.
(2002) for $n=1$ and $\gamma=1$. The form (\ref{InLargeX}) is
similar to the form of (\ref{equ15}) and (\ref{equ16}) in the
overall scaling. The integration constant $H$ in the invariant
form is an analogy of the `velocity parameter' $B$, while the
`mass parameter' $A$ is already absorbed into the parameter $D$ in
the invariant form. To make this asymptotic solution applicable,
we should require $H=0$ in cases of $n<1$ and thus $\gamma>1$. The
boundary conditions at $x\rightarrow +\infty$ correspond to the
initial conditions of the fluid. Therefore, the initial density
scales as $r^{-l}$. More precisely, this new self-similar
transformation requires the initial density profile to scale as
$r^{-2/n}$, identical with the scaling we obtain in the main text
using self-similar transformation (\ref{equ6}). For $x\rightarrow
0^+$, the asymptotic solution reads
\begin{equation}
v\sim 2x/3\ , \qquad R\sim Ix^l\ , \qquad\hbox{ as }
x\rightarrow0^+\ , \label{InSmallx}
\end{equation}
where $I>0$ is an arbitrary parameter for a positive mass density.
This $v$ expression of asymptotic solution (\ref{InSmallx}) is
independent of $l$, different from the $v$ expression (35) in
\citet{Shu}, but it is similar to asymptotic solution
(\ref{equ13}) in the main text; the $R$ expression of our
asymptotic solution is similar to the $R$ expression (35) of
\citet{Shu}, and the parameter $I$ here is proportional to the
central reduced mass density $\alpha_0$. In the isothermal case
with $n=1$ and $\gamma=1$, relation (\ref{M2}) is satisfied
automatically and parameter $l$ is thus arbitrary. Here lies a
fundamental difference between the conventional polytropic case
and the isothermal case.


\label{lastpage}

\end{document}